\documentclass[aps,prd,superscriptaddress,nofootinbib,preprint]{revtex4-1}
\pdfoutput=1
\usepackage[utf8]{inputenc}
\usepackage[english]{babel}
\usepackage{amsmath}
\usepackage{subfigure}
\usepackage{amsfonts}
\usepackage{amssymb}
\usepackage{color}
\usepackage{cancel}
\usepackage{graphicx}

\begin{document}

\title{Signatures of Supersymmetry  in Neutrino Telescopes}

\author{P. S. Bhupal Dev}

\affiliation{Department of Physics and McDonnell Center for the Space Sciences, Washington University, St. Louis, MO 63130, USA}
\affiliation{Theoretical Physics Department, Fermi National Accelerator Laboratory, P.O. Box 500, Batavia,
IL 60510, USA.}
\begin{abstract}
We review the prospects of probing $R$-parity violating Supersymmetry (RPV SUSY) at neutrino telescopes using some of the highest energy particles given to us by Nature. The presence of RPV interactions involving ultra-high energy neutrinos with Earth-matter can lead to resonant production of TeV-scale SUSY partners of the SM quarks and leptons (squarks and sleptons), thereby giving rise to potentially anomalous behavior in the event spectrum observed by  large-volume neutrino detectors, such as IceCube, as well as balloon-borne cosmic ray experiments, such as ANITA. Using the ultra-high energy 
neutrino events observed recently at IceCube, with the fact that for a 
given power-law flux of astrophysical neutrinos, there is no statistically 
significant deviation in the current data from the Standard Model 
expectations, we derive robust upper limits on the RPV couplings as a 
function of the resonantly-produced squark mass, independent of the 
other unknown model parameters, as long as the squarks decay dominantly to two-body final states involving leptons and quarks through the RPV couplings. Also, we discuss RPV SUSY interpretations of the recent anomalous, upward-going EeV air showers observed at ANITA, in terms of long-lived charged or neutral next-to-lightest SUSY particles.  
\end{abstract}

\maketitle

\section{Introduction} \label{sec:1}
The Standard Model (SM) of particle physics has worked beautifully to predict what experiments have shown so far about the basic building blocks of matter~\cite{Tanabashi:2018oca}. However, physicists have long recognized that it cannot be the ultimate theory of Nature. Experimental evidence in support of this include the observation of nonzero neutrino mass, dark matter (DM), dark energy, and matter-antimatter asymmetry, none of which the SM can explain. Among the theoretical problems the SM fails to address are the quadratic divergence of the Higgs mass (the so-called hierarchy problem), the adhoc choice of the gauge groups and particle representations, unification of gauge couplings, inclusion of gravity, and the origin of electroweak symmetry breaking. While these arguments all point to some beyond the SM (BSM) physics, without further assumptions, they do not point decisively to the energy scale at which it should occur. 

However, given the observed Higgs boson mass of 125 GeV, the quadratically divergent corrections in the SM imply that unless the new physics scale $\Lambda$ is close to the TeV scale, we need unnatural fine-tuning of parameters to keep the Higgs mass at its observed value. For instance, if $\Lambda \sim 10^{16}$ GeV, i.e. at the Grand Unified Theory (GUT) scale, then the mass-squared parameter in the SM Lagrangian will have to be fine-tuned to 1 part in $10^{26}$ to maintain a physical Higgs mass at 125 GeV. Although not logically impossible, such extreme fine-tuning is usually thought to be symptomatic of a deeper issue, often dubbed as the naturalness problem. If taken seriously, this leads to the inevitable conclusion that there must be new degrees of freedom around TeV scale.

Supersymmetry (SUSY), which predicts the existence of a partner differing in spin by $\frac{1}{2}$  to every known particle is widely considered as the most attractive solution to the naturalness and hierarchy problems. Specifically, due to the new symmetry between bosons and fermions, the quadratically  divergent loop contributions of the SM particles to the Higgs mass squared are exactly canceled by the corresponding loop contributions of their SUSY partners. There are numerous other theoretical and phenomenological motivations for SUSY, such as (i) the most general symmetries of the $S$-matrix in quantum field theory are a direct product of the super-Poincar\'e group, which includes SUSY, with the internal symmetry group; (ii) gravity is naturally incorporated if SUSY is made local, as in supergravity models; (iii) SUSY is an essential ingredient of superstring theories, which are by far the best candidates for a consistent quantum theory of gravity; (iv)  the SUSY particle spectrum leads to the remarkable unification of the strong and electroweak gauge couplings at an energy scale of $M_{\rm GUT}\sim 2\times 10^{16}$ GeV, which is consistent with proton decay constraints; (v) the lightest SUSY particle (LSP), if electrically and color neutral, makes an excellent candidate for DM; and (vi) the observed Higgs boson mass is consistent with the upper limit of $\lesssim 135$ GeV on the lightest Higgs boson in the minimal SUSY SM (MSSM).  All these arguments in favor of SUSY make it arguably the strongest contender for BSM physics~\cite{Haber:1984rc, Baer:2006rs}. 

However, the lack of evidence for superpartners in the Large Hadron Collider (LHC) data so far~\cite{atlas-susy, cms-susy} has forced the simplest SUSY scenarios toward regions of parameter space unnatural for the Higgs sector~\cite{Papucci:2011wy, Hall:2011aa, Brust:2011tb, Buckley:2016kvr, Ahmed:2017jxl}. A simple way to preserve the Higgs naturalness by evading the current experimental constraints is by allowing $R$-parity Violation (RPV)~\cite{Barbier:2004ez} in the production and decays of superpartners at the LHC. Apart from significantly lowering the collider bounds on the SUSY spectrum~\cite{Carpenter:2006hs, Dreiner:2012np, Allanach:2012vj, Dreiner:2012wm, Asano:2012gj, Graham:2014vya, Monteux:2016gag, Dercks:2017lfq, Guo:2018hbv, Bansal:2018dge}, RPV SUSY implies the violation of baryon and/or lepton numbers, which has important phenomenological consequences~\cite{Barbier:2004ez}. For instance, one can automatically generate non-zero neutrino masses and mixing~\cite{Hall:1983id,  Ellis:1984gi, Dawson:1985vr, Joshipura:1994ib, Hempfling:1995wj, Roy:1996bua, Mukhopadhyaya:1998xj, 
 Bhattacharyya:1999tv, Davidson:2000ne, Diaz:2003as, Dreiner:2011ft, Peinado:2012tp} at either tree or one-loop level without introducing any extra particles beyond the MSSM field content. Similarly, the presence of $\Delta L\neq 0$ RPV vertices can lead to observable neutrinoless double beta decay ($0\nu\beta\beta$)~\cite{Mohapatra:1986su, Vergados:1986td, Hirsch:1995zi, Babu:1995vh, Pas:1998nn, Faessler:2007nz, Allanach:2009xx, Rodejohann:2011mu}, as well as successful baryogenesis~\cite{Cline:1990bw, Masiero:1992bv, Sarkar:1996sn, Dolgov:2006ay, Kohri:2009ka, Cui:2012jh, Sorbello:2013xwa, Arcadi:2015ffa, Farina:2016ndq, Pierce:2019ozl}. Moreover, RPV scenarios also provide a compelling solution to some recent anomalies, such as the muon anomalous magnetic moment~\cite{Kim:2001se, Bhattacharyya:2009hb, Hundi:2011si, Chakraborty:2015bsk}, lepton flavor universality violation in semileptonic $B$-meson decays~\cite{Deshpande:2012rr, Biswas:2014gga, Huang:2015vpt, Zhu:2016xdg, Deshpand:2016cpw, Altmannshofer:2017poe, Das:2017kfo, Earl:2018snx, Sheng:2018ylg, Trifinopoulos:2018rna, Hu:2018lmk} and anomalous upgoing events at ANITA~\cite{Collins:2018jpg}.  
If any of these anomalies becomes statistically significant, one should consider RPV SUSY as a strong candidate for the underlying new physics. Hence, it is of paramount importance to find complementary ways at as many different energy scales as possible to test this scenario. 

In this review, we discuss one such possibility, namely, testing the RPV SUSY scenario using ultra-high energy (UHE) neutrinos at neutrino telescopes, such as IceCube and KM3NeT. The basic idea is that the presence of RPV SUSY interactions involving UHE neutrinos can lead to resonant production of TeV-scale sfermions inside the Earth~\cite{Carena:1998gd, Dev:2016uxj, Collins:2018jpg}, thereby giving rise to distinct features in the detected neutrino spectrum at neutrino telescopes.\footnote{Even in the absence of RPV, one could consider pair production of sparticles in neutrino-nucleon interactions~\cite{Albuquerque:2003mi}. In this case, the relatively small cross-section for the production of sparticles, as compared to the SM neutrino-nucleon cross section, can be partially compensated for, provided one of the sparticles is very long lived~\cite{Albuquerque:2003mi, Ahlers:2006pf, Albuquerque:2006am, Ando:2007ds, Canadas:2008ey,  Meade:2009mu, Connolly:2018ewv}.  We will briefly discuss this possibility in Section~\ref{sec:5}.}  Note that in the SM, the only observable resonance for UHE neutrinos is the Glashow resonance~\cite{Glashow:1960zz}: $\bar{\nu}_ee^-\to W^-$, which occurs at incoming electron antineutrino energy $E_\nu=\frac{m_W^2}{2m_e}=6.3$ PeV. In contrast, the squark/slepton resonance can occur at a different incoming neutrino energy, depending on the mass of the SUSY particles. Since no such resonance feature has been observed in the IceCube data so far~\cite{Aartsen:2017mau} (except for a mild $2\sigma$-level discrepancy around 100 TeV which could very well be an artifact of the fitting procedure or astrophysical flux uncertainty), one can derive robust upper limits on the
RPV couplings as a function of the resonantly-produced sfermion mass, largely independent of the other unknown model parameters~\cite{Dev:2016uxj}.  With more statistics, we expect these limits to be
comparable/complementary to the existing limits from direct collider searches and other low-energy processes. More importantly, if IceCube does observe a statistically significant resonance-like excess feature in the future, it might point to a preferred range of the relevant RPV coupling and sfermion mass that could be looked for more vigorously in the LHC data.

The rest of this article is organized as follows: in Section~\ref{sec:2}, we briefly discuss the RPV interactions in the MSSM. In Section~\ref{sec:3}, we focus on the UHE neutrino interactions involving trilinear RPV. In Section~\ref{sec:4}, we calculate the event rate at IceCube and derive constraints on the $\lambda'$-type RPV couplings as a function of the squark mass. In Sections~\ref{sec:5} and \ref{sec:6}, we consider the possibility of long-lived charged and neutral SUSY particles respectively, and implications for neutrino telescopes, as well as balloon experiments. We conclude in Section~\ref{sec:7}. 
 
\section{RPV Interactions} \label{sec:2}
In the SUSY extension of the SM, all the fields, both fermionic and bosonic, get promoted to become chiral superfields which have both a bosonic and a fermionic part. The bosonic partners of known SM fermions are called sfermions (i.e. squarks and sleptons), whereas the fermionic partners of the Higgs fields are called Higgsinos, and those of the SM gauge fields are called gauginos (i.e., gluino, bino, wino, photino).

$R$-parity is associated with a $Z_2$ subgroup of the group of continuous $U(1)$ $R$-symmetry transformations acting on the gauge superfields and the two chiral doublet Higgs superfields $H_u$ and $H_d$ responsible for electroweak symmetry breaking~\cite{Fayet:1974pd}. $R$-parity appears as the discrete remnant of this continuous $U(1)_R$ symmetry that must be broken to allow for the gravitino (the superpartner of graviton in supergravity models) and gluinos to acquire masses. There is an intimate connection between $R$-parity and baryon ($B$) and lepton ($L$) number conservation laws~\cite{Farrar:1978xj}, which is made explicit by the following expression for the $R$-parity of a particle with spin $S$: 
\begin{align}
(-1)^R \ = \ (-1)^{3(B-L)+2S} \, .
\end{align}
Under this symmetry, all the SM fields are even, while all the superpartner fields are odd. Thus, the conservation of $R$-parity implies that all superpartner fields must be produced in pairs and must decay to states containing at least one superpartner field. This also implies that the LSP is absolutely stable, which has two cosmological implications: (i) it cannot be electrically charged, as it is cosmologically disfavored~\cite{DeRujula:1989fe, Dimopoulos:1989hk, Pospelov:2006sc}, and (ii) if electrically and color neutral, it is a good candidate for DM~\cite{Jungman:1995df}. 

The MSSM superpotential invariant under the SM gauge group $SU(3)_c\times SU(2)_L\times U(1)_Y$ that preserves $R$-parity is written as 
\begin{align}
W_{\rm MSSM} \ = \ \mu H_uH_d+h_uQ H_uU^c+h_dQH_dD^c+h_eLH_dE^c \, ,
\label{eq:sup}
\end{align}  
where $L\ni (\nu, e)_L$ and $Q \ni (u, d)_L$ are the $SU(2)_L$-doublet and $U^c, D^c, E^c$ are the $SU(2)_L$-singlet chiral superfields, respectively. Since SUSY is not an exact symmetry of nature, one must also add the SUSY breaking terms, which consist of soft mass terms for superpartners and trilinear $A$-terms of the type in Eq.~\eqref{eq:sup} but with the scalar components of the superfields and mass terms for gauginos, all of which still respect the gauge symmetry. 

As mentioned in Section~\ref{sec:1}, in light of the current experimental results, the case for $R$-parity conservation (RPC) in effective low-energy SUSY is less compelling than RPV. We have a few additional points in favor of RPV:  
\begin{itemize}
\item $R$-parity by itself is not enough to protect against dangerous proton decay operators~\cite{Nath:2006ut}, as RPC but non-renormalizable superpotential interactions of the general form $U^cU^cD^cE^c$ can give rise to proton decay. So the standard proton-stability motivation for $R$-parity is actually gone. 

\item Although RPV interactions spoil the stability of the DM, it is still possible to have the LSP long-lived enough to play the role of DM, e.g. gravitino because of its Planck-suppressed gravitational interactions~\cite{Takayama:2000uz, LopezFogliani:2005yw, Buchmuller:2007ui, Monteux:2014tia}. Another possibility is to have a different symmetry than the $R$-parity, under which the SM is inert. 
  
\item Introducing RPV at low scale still preserves the gauge coupling unification~\cite{Altmannshofer:2017poe} -- a hallmark prediction of SUSY. In addition, RPV can be made compatible with the unification of quarks and leptons in GUT models~\cite{DiLuzio:2013ysa, Bajc:2015zja}. In fact, large RPV couplings are favorable to realize the $b-\tau$ Yukawa unification~\cite{Hu:2018lmk}.  

\item RPV is mandatory in minimal theories that extend MSSM to include local $B-L$ symmetry~\cite{Mohapatra:1986su, Martin:1992mq, Mohapatra:2015fua}. Since the nonzero neutrino masses seem to suggest that the underlying BSM physics must include a broken local $B-L$ symmetry, the fact that its supersymmetrization leads to RPV makes it even more compelling.       
\end{itemize}  

There are two classes of RPV models: (i) spontaneous breaking by giving vacuum expectation values to neutral superpartners, such as sneutrino~\cite{Aulakh:1982yn}, or (ii) explicit breaking by adding RPV terms in the superpotential~\eqref{eq:sup}~\cite{Hall:1983id}. In this review, we will focus on the latter case, where the explicit RPV terms in MSSM can be written as 
\begin{align}
W_{\rm RPV} \ = \ & \mu_i L_i H_u + \frac{1}{2}\lambda_{ijk}L_iL_jE_k^c+\lambda'_{ijk} L_i Q_j D_k^c 
+\frac{1}{2}\lambda''_{ijk}U_i^c D_j^c D_k^c \, ,
\label{supRPV}
\end{align}
where $i,j,k=1,2,3$ are the family indices and we have suppressed all gauge indices for brevity. Note that $SU(2)_L$ and $SU(3)_c$ gauge invariance enforce antisymmetry of the $\lambda_{ijk}$- and $\lambda''_{ijk}$-couplings with respect to their first and last two indices, respectively. 

Since we are mostly interested in the UHE neutrino interactions with nucleons, we will only focus on the $\lambda'_{ijk}$-couplings. Any of these 27 new dimensionless complex parameters can lead to resonant production of TeV-scale squarks at IceCube energies~\cite{Carena:1998gd, Dev:2016uxj}, thereby making a potentially significant contribution to the UHE neutrino events.\footnote{Even without SUSY, similar resonance features in neutrino-nucleon interactions can also occur in models with TeV-scale leptoquarks~\cite{Robinett:1987ym, Doncheski:1997it, Anchordoqui:2006wc, Alikhanov:2013fda, Barger:2013pla, Dutta:2015dka, Dey:2015eaa, Dey:2017ede, Becirevic:2018uab}. However, there are subtle differences between scalar leptoquark and RPV SUSY models, e.g. due to the presence of additional decay channels and chiral mixing between squarks in the RPV case.} The $\lambda$-couplings in Eq.~\eqref{supRPV} can give rise to resonant production of selectrons from neutrino interactions with electrons, reminiscent of the Glashow resonance~\cite{Glashow:1960zz} in the SM; however, for TeV-scale selectrons, their contribution to the total number of IceCube events is negligible in the energy range considered here and we will comment on this possibility later in the context of ANITA events. Note that with non-zero $\lambda'$-couplings, we need to explicitly forbid the $\lambda''$-terms, e.g.~by imposing baryon triality~\cite{Ibanez:1991hv}, to avoid rapid proton decay, as the product $\lambda'\lambda''$ is constrained to be $\lesssim 10^{-24}$ from proton lifetime bounds~\cite{Smirnov:1996bg}. We also ignore the bilinear terms in Eq.~\eqref{supRPV}, since they do not give rise to the resonance feature exploited here, although they  could lead to other distinct signatures (e.g. triple bang) relevant for future multi-km$^3$ neutrino telescopes~\cite{Hirsch:2007kx}.

\section{Neutrino Interactions involving Trilinear RPV} \label{sec:3} 
\begin{figure}[t]
\centering
\includegraphics[width=0.6\textwidth]{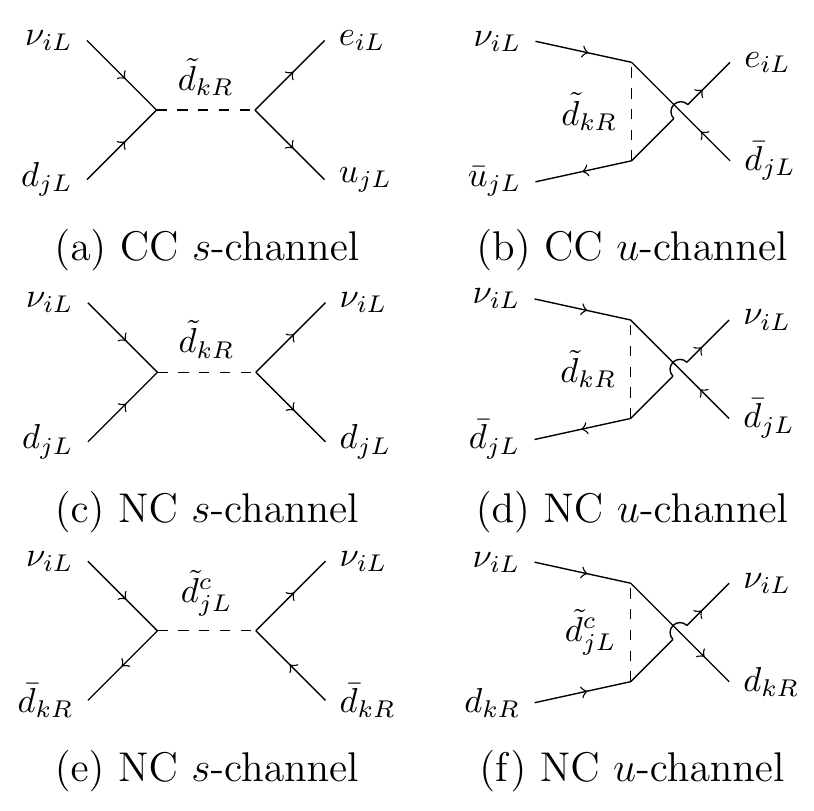}
\caption{Representative Feynman diagrams for the CC and NC contributions to the neutrino-nucleon interactions induced by the $\lambda'_{ijk}$-terms in Eq.~\eqref{laglamp}. }
\label{fig1}
\end{figure}

We start with the $\lambda'$-part of the RPV Lagrangian, after expanding the superpotential~\eqref{supRPV} in terms of the superfield components: 
\begin{align}
& {\cal L}_{LQD}  \ =  \   \lambda'_{ijk} \bigg[\tilde{\nu}_{iL} \bar{d}_{kR} d_{jL}  
+ \tilde{d}_{jL}  \bar{d}_{kR} \nu_{iL} 
+ \tilde{d}_{kR}^* \bar{\nu}_{iL}^c d_{jL} 
\nonumber \\
& \qquad \qquad 
- \tilde{e}_{iL} \bar{d}_{kR}  u_{jL} 
  - \tilde{u}_{jL}  \bar{d}_{kR}  e_{iL}  
- \tilde{d}_{kR}^* \bar{e}_{iL}^c u_{jL}  \bigg] +{\rm H.c.} 
\label{laglamp}
\end{align}
At the IceCube, these interactions will contribute to both charged-current (CC) and neutral-current (NC) processes mediated by either $s$-channel or $u$-channel exchange of a down-type squark, as shown in Fig.~\ref{fig1}. 

The $s$-channel processes in Figs.~\ref{fig1}(a) and (c) involve valence quarks, thus giving the dominant contributions to the (anti)neutrino--nucleon cross sections, provided the right-handed (RH) down-type squarks are produced resonantly. Similarly, the $s$-channel process in Fig.~\ref{fig1}(e) mediated by a left-handed (LH) down-type squark can also give a resonant enhancement to the (anti)neutrino--nucleon cross section. Here we have implicitly assumed that the RPC squark decays to a quark and a gluino, neutralino or chargino are suppressed~\cite{Butterworth:1992tc}, compared to the RPV decays induced by Eq.~\eqref{laglamp}. This is the case, for instance, in the region of RPV MSSM parameter space, where the gaugino masses $M_1$, $M_2$, as well as the $\mu$-term, are larger than the squark masses, thus kinematically forbidding the two body RPC decays of squarks. The 3-body RPC decays via virtual gauginos will in general be smaller compared to the 2-body decays through RPV couplings, as considered here. 

On the other hand, the contributions from the $u$-channel processes in Fig.~\ref{fig1}(b), (d) and (f) are much smaller, since they do not have a resonant enhancement, and in addition, for (b) and (d),  due to the sea quark involvement. Moreover, the RPV contributions are the largest for the first generation quarks, which are the predominant constituents of the nucleon, and to some extent, for the second and third generation quarks, which become increasingly important at higher energies. At the same time, the existing indirect constraints from precision measurements in various low-energy processes are the strongest for the couplings involving the first generation and weaker for the second and third generations~\cite{Barbier:2004ez, Allanach:1999ic, Kao:2009fg, Dreiner:2012mx, Domingo:2018qfg}. For the SM CC and NC interactions~\cite{Gandhi:1995tf, CooperSarkar:2011pa}, which must be included in the total neutrino-nucleon cross section giving rise to the IceCube events, we take into account all valence and sea quark contributions.  

The total differential cross section for the neutrino-nucleon interactions, written in terms of the Bjorken scaling variables $x=Q^2/2m_NE'_\nu$ and $y=E'_\nu/E_\nu$, is given by 
\begin{align}
\frac{d^2\sigma}{dxdy} \ = & \ \frac{m_N E_\nu}{16\pi} \sum_f \bigg[xf(x,Q^2)|a_f|^2 
+x\bar{f}(x,Q^2)|b_f|^2(1-y)^2\bigg] \; , \label{cross}
\end{align}
where $m_N=(m_p+m_n)/2$ is the average mass of the proton and neutron for an isoscalar nucleon, $-Q^2$ is the invariant momentum transfer between the incident neutrino and 
outgoing lepton, $E_\nu$ is the incoming neutrino energy, $E'_\nu=E_\nu-E_\ell$ is the energy loss in the laboratory frame, $E_\ell$ is the energy of the outgoing lepton, and $f(x,Q^2),\bar{f}(x,Q^2)$ are the parton distribution functions (PDFs) within the proton for $f$-quark and anti $f$-quark, respectively. For the CC processes shown in Figs.~\ref{fig1}(a) and (b), induced by an incoming neutrino of flavor $i$, the only non-trivial coefficients in Eq.~\eqref{cross} are respectively~\cite{Carena:1998gd}  
\begin{align}
a_{d_j}^{\rm CC} \ & = \ \frac{g^2}{Q^2+m_W^2} - \sum_k \frac{|\lambda'_{ijk}|^2}{xs-m^2_{\tilde{d}_{kR}}+im_{\tilde{d}_{kR}}\Gamma_{\tilde{d}_{kR}}}, \label{acc} \\
b_{\bar{u}_j}^{\rm CC} \ & = \ \frac{g^2}{Q^2+m_W^2} - \sum_k \frac{|\lambda'_{ijk}|^2}{Q^2-xs-m^2_{\tilde{d}_{kR}}}, \label{bcc}
\end{align}
where $s=2m_N E_\nu$ is the square of the center-of-mass energy and $g$ is the $SU(2)_L$ gauge coupling, and $m_W$ is the $W$-boson mass. It is obvious to see that in Eqs.~\eqref{acc} and \eqref{bcc}, the first term on the right-hand side is the SM CC contribution, whereas the second term is the RPV contribution. So for all SM CC processes involving $\bar{d}$ and $u$-type quarks, which do not have interference with the RPV processes, the coefficients in Eq.~\eqref{cross} are simply obtained by putting $\lambda'_{ijk}=0$ in Eqs.~\eqref{acc} and \eqref{bcc}. 

Similarly, for the NC processes shown in Figs.~\ref{fig1}(c)--(f), the only non-trivial coefficients are~\cite{Carena:1998gd}
\begin{align}
a_{d_j}^{\rm NC} \ & = \ \frac{g^2}{1-x_w}\frac{L_d}{Q^2+m_Z^2} - \sum_k \frac{|\lambda'_{ijk}|^2}{xs-m^2_{\tilde{d}_{kR}}+im_{\tilde{d}_{kR}}\Gamma_{\tilde{d}_{kR}}}, \label{anc} \\
b_{d_j}^{\rm NC} \ & = \  \frac{g^2}{1-x_w}\frac{R_d}{Q^2+m_Z^2} - \sum_k \frac{|\lambda'_{ijk}|^2}{Q^2-xs-m^2_{\tilde{d}_{kL}}}, \label{bnc}
\end{align}
where $L_d=-(1/2)+(1/3)x_w$ and $R_d=(1/3)x_w$ are the chiral couplings, $x_w\equiv \sin^2\theta_w$ is the weak mixing angle parameter and $m_Z$ is the $Z$-boson mass. For all SM NC processes involving $u$-type quarks, which do not have interference with the RPV processes, the coefficients in Eq.~\eqref{cross} are simply obtained by putting $\lambda'_{ijk}=0$ in Eqs.~\eqref{anc} and \eqref{bnc} and replacing $L_d\to L_u= (1/2)-(2/3)x_w$, $R_d\to R_u=-(2/3)x_w$.
For neutrino-antiquark interactions, the coefficients for the NC processes can be obtained simply by crossing symmetry, i.e.~$a_f\leftrightarrow b_f,\ xs\leftrightarrow Q^2-xs$. Similarly, for antineutrino-nucleon interactions, we can just replace the PDFs $f \leftrightarrow \bar{f}$ in Eq.~\eqref{cross}. 

Note the Breit-Wigner resonance form of Eqs.~\eqref{acc} and \eqref{anc}, which is regulated by the RH down-type squark width 
\begin{align}
\Gamma_{\tilde{d}_{kR}} \ \simeq \ \frac{m_{\tilde{d}_{kR}}}{8\pi}\sum_{ij} |\lambda'_{ijk}|^2,
\label{decaydR}
\end{align}
 assuming that the only dominant decay modes are $\tilde{d}_{kR}\to \nu_{iL}d_{jL}$ (NC) and $\tilde{d}_{kR} \to e_{iL}u_{jL}$ (CC), and the masses of the final state fermions in these 2-body decays are negligible compared to the parent squark mass. For the LH down-type squark, $\Gamma_{\tilde{d}_{kL}}=\Gamma_{\tilde{d}_{kR}}/2$, since $\tilde{d}_{kL}\to \nu_{iL}d_{jR}$ is the only available decay mode. The resonance condition is satisfied for the incoming energy $E_\nu=m^2_{\tilde d_{kR}}/2m_Nx$, but due to the spread in the initial quark momentum fraction $x\in [0,1]$, the resonance peak will be broadened and shifted above the threshold energy $E_\nu^{\rm th}=m^2_{\tilde d_{kR}}/2m_N$. Nevertheless, given that the maximum contribution to the cross section happens around $x\sim 10^{-3}-10^{-4}$, one can immediately infer that for $m_{\tilde d_{kR}}\in$ [100  GeV, 2 TeV], $E_\nu^{\rm th}$ is in the multi TeV--PeV range, and hence, can be probed by the available IceCube data.  Interestingly, this is exactly the mass range currently being probed at the LHC, and therefore, the neutrino telescope probes allow for an independent, complementary test of the SUSY parameter space.
 
Now let us consider the $\lambda$-couplings in Eq.~\eqref{supRPV}, which lead to the $LLE$-type RPV Lagrangian
\begin{align}
{\cal L}_{LLE} \ = & \ \frac{1}{2}\lambda_{ijk}\bigg[ \tilde{\nu}_{iL}\bar{d}_{kR}d_{jL}+\tilde{e}_{iL}\bar{e}_{kR}\nu_{jL}+\tilde{e}^*_{kR}\bar{\nu}^c_{iL}e_{jL}
-(i\leftrightarrow j)\bigg]+{\rm H.c.}
\label{eq:LLE}
\end{align}
With these interactions, we can have new contributions to the (anti)neutrino-electron scattering inside Earth. In particular, given enough energy of the incoming (anti)neutrino, this will lead to the resonant production of an LH slepton through the second term in Eq.~\eqref{eq:LLE}, and similarly an RH slepton through the third term in Eq.~\eqref{eq:LLE}. For an incoming neutrino energy $E_\nu$, the slepton mass at which the resonance occurs is simply $m_{\widetilde{e}_i}=\sqrt{s}=\sqrt{2E_\nu m_e}$, where $s$ is the center-of-mass energy~\cite{Carena:1998gd, Dev:2016uxj}. This is reminiscent of the Glashow resonance in the SM, where an on-shell $W$ boson is produced from the $\bar{\nu}_e-e$ scattering with an initial neutrino energy of $E_\nu=m_W^2/2m_e=6.3$ PeV~\cite{Glashow:1960zz}.   However, the corresponding threshold energy $E_\nu^{\rm th}=m^2_{\tilde{e_k}}/2m_e$ is beyond 10 PeV for selectron masses above 100 GeV or so. Since smaller selectron masses are excluded from the LEP data~\cite{Abbiendi:2003rn, Abdallah:2003xc}, we cannot probe the $\lambda_{ijk}$-couplings with the current IceCube data. Nevertheless, if future data reports any events beyond 10 PeV, the $LLE$-type RPV scenario could in principle provide a viable explanation, given the fact that it would be  difficult to explain those events within the SM and with an unbroken power-law flux, without having a significantly larger number of events in all the preceding lower-energy bins.    We will come back to Eq.~\eqref{eq:LLE} in Section~\ref{sec:6} in a different context.

\section{Event Rate} \label{sec:4}
To give an illustration of the power of neutrino telescope in testing the RPV SUSY parameter space, let us calculate the event rate for IceCube including the RPV effect. In general, the expected number of high-energy starting events (HESE) in a given deposited energy bin at IceCube due to the modified cross section~\eqref{cross} can be estimated as
\begin{align}
N_{\rm bin} \ = \ TN_A\int_{E^{\rm bin}_{\rm min}}^{E^{\rm bin}_{\rm max}}dE_{\rm dep} \int_0^1 dy \: V_{\rm eff} \:  \Phi \: \Omega \: \frac{d\sigma}{dy} \, ,
\label{event}
\end{align}
where $T$ is the exposure time, $N_A$ is the Avogadro number, $E_{\rm dep}(E_\nu)$ is the electromagnetic-equivalent deposited energy which is always smaller than the incoming neutrino energy $E_\nu$ in the laboratory frame by a factor depending on $E_\nu$ and the type of interaction (CC or NC) and lepton flavor,  $V_{\rm eff}(E_\nu)$ is the effective target volume of the detector, $\Phi(E_\nu)$ is the incident neutrino flux, $\Omega(E_\nu)$ is the effective solid angle of coverage, and we have integrated the differential cross section in Eq.~\eqref{cross} over $x\in [0,1]$, including both neutrino and antineutrino initial states with all flavors. For more details of the event rate calculation, see e.g. Refs.~\cite{Laha:2013eev, Chen:2013dza, Aartsen:2013vja, Aartsen:2013jdh, Vincent:2016nut, Palladino:2018evm, Sui:2018bbh}. 

We calculate the predicted number of events with and without RPV interactions for the publicly available 6-year IceCube HESE data corresponding to $T=2078$ days. Here we have assumed the IceCube best-fit~\cite{Aartsen:2017mau} for a single-component unbroken astrophysical power-law flux $\Phi(E_\nu) = 2.5\times 10^{-18}(E_\nu/100~{\rm TeV})^{-2.9}~{\rm GeV}^{-1}\: {\rm cm}^{-2}{\rm s}^{-1}{\rm sr}^{-1}$ with a standard (1:1:1) flavor composition ratio on Earth, and have used the NNPDF2.3 leading order PDF sets~\cite{Ball:2012cx} for the cross section calculations.\footnote{The PDF uncertainties on the total cross-section are at most at 5\% level~\cite{Chen:2013dza} for the energy range of interest. The flux uncertainties, on the other hand, are currently at the level of 15\%~\cite{Aartsen:2017mau}. In addition, there is a tension between the HESE and throughgoing muon samples, which cannot be resolved by the given single-component unbroken power-law HESE fit and might require some BSM explanation by itself~\cite{Sui:2018bbh}. } 
The expected background due to atmospheric neutrinos and muons are also taken from the IceCube analysis~\cite{Aartsen:2017mau}. 
Note that 
atmospheric $\nu_e$ events will also get modified due to 
nonzero RPV couplings $|\lambda'_{11k}|$. However, since the atmospheric 
background is dominated by the $\nu_\mu$-induced events and the event rate 
for atmospheric $\nu_e$ is much smaller,  we can 
safely ignore the $|\lambda'_{11k}|$ effects on the background and just 
assume it to be basically the same as in the SM case. Also one might wonder 
whether the source flavor composition and flux of the neutrinos could be 
modified due to the new RPV interactions. However, this effect is expected 
to be small for the values of squark masses and RPV couplings considered here, 
since the SM weak interactions with strength 
$G_F$ (the Fermi coupling constant) will be dominant over the RPV 
interactions with relative strength of $|\lambda'_{ijk}|^2/m_{\tilde d_k}^2$. 
The RPV effect at the IceCube detector could get enhanced {\it only} due to the resonant 
production of squarks in a conducive range of the incoming neutrino energy. Thus, adding the flux uncertainty in our analysis will equally affect the events due to both SM and RPV interactions, without changing the relative enhancement of the events in presence of the RPV interactions with respect to the SM prediction. This justifies our use of the IceCube best-fit value for the flux.   

\begin{figure}[t]
\centering
\includegraphics[width=0.5\textwidth]{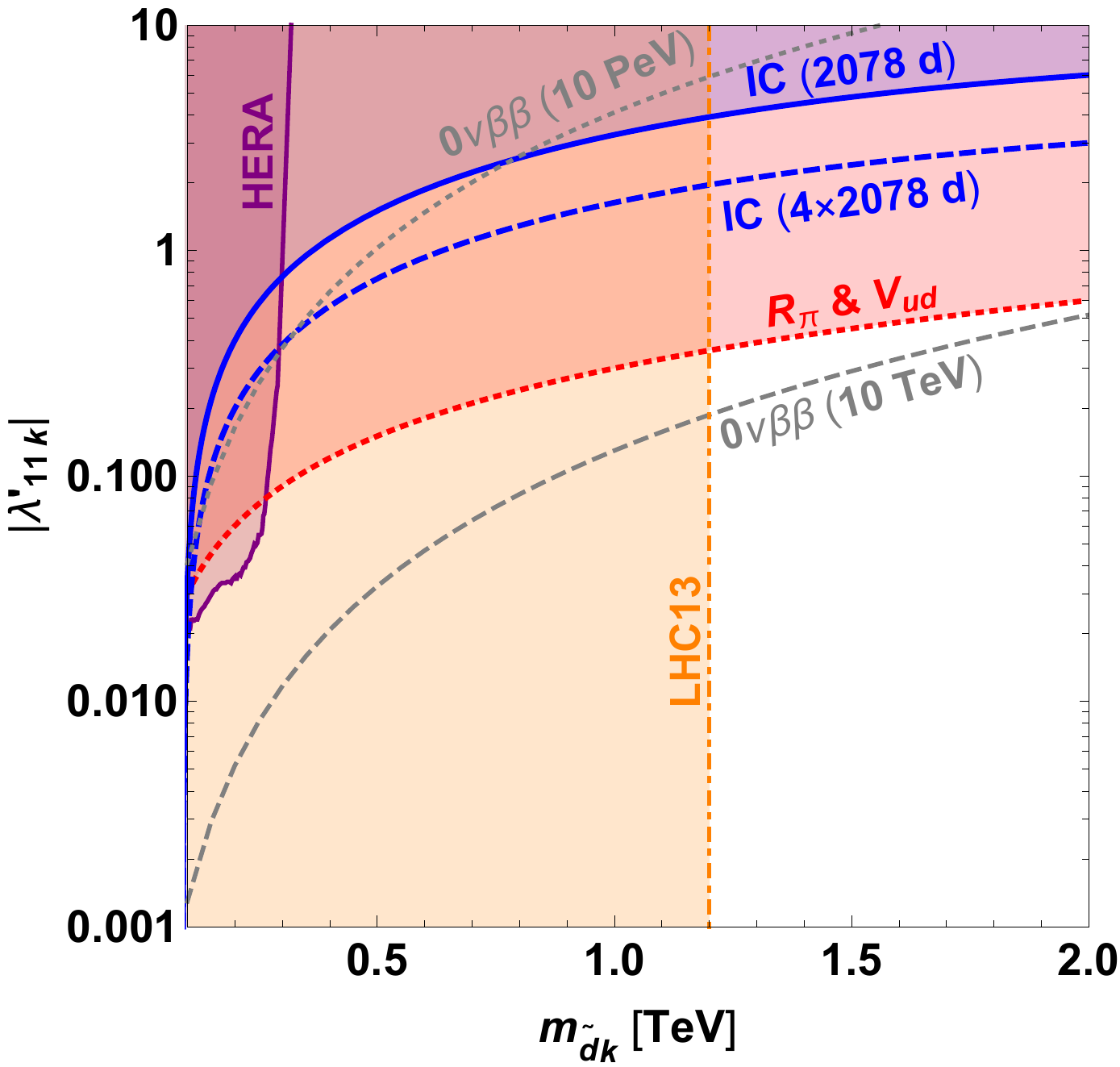}  
\includegraphics[width=0.48\textwidth]{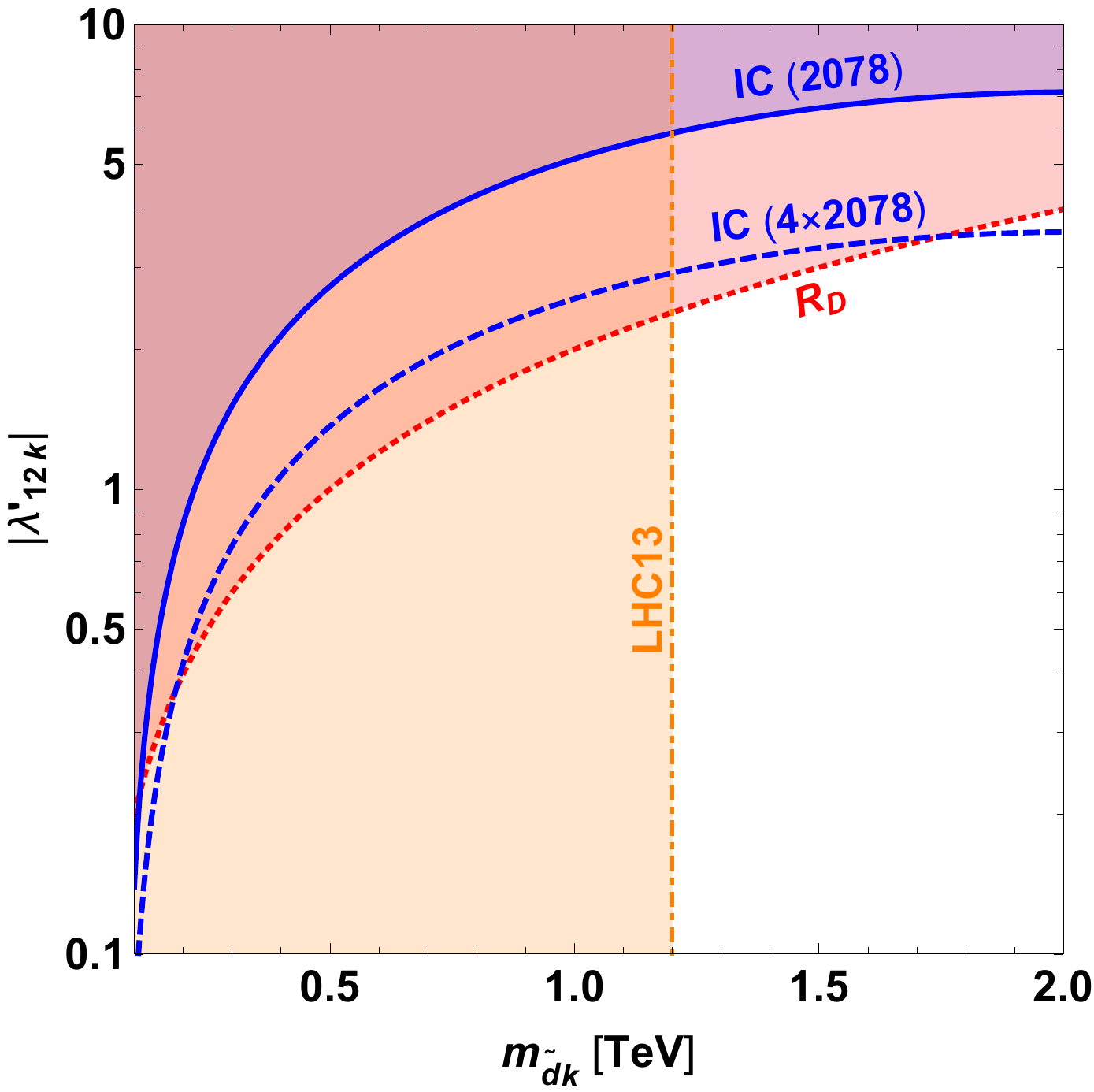} \\
\caption{$1\sigma$ upper limits (blue, solid) on the RPV couplings $|\lambda'_{1jk}|$ (with $j=1,2$) as a function of the down-type squark mass $m_{\tilde{d}_k}$ from the 8-year IceCube HESE data and projected limits (blue, dashed) obtained by scaling the exposure time by a factor of 4. For comparison, we also show the $2\sigma$ indirect limits from lepton universality in meson decays (red, dotted) and the 95\% CL direct limits from a scalar leptoquark search at the 13 TeV LHC (orange, dot-dashed). In the left panel, we additionally show the 95\% CL direct search limit (magenta, solid) from $e^-p$ collisions at HERA, as well as the 90\% CL $0\nu\beta\beta$ limits for two benchmark values of 10 TeV (gray, dashed) and 10 PeV (gray, dotted) for the gaugino masses, while keeping all the relevant sfermion masses fixed at $m_{\tilde{d}_k}$. This figure is an updated version of Fig.4 in~\cite{Dev:2016uxj} with the new IceCube,  $0\nu\beta\beta$ and LHC results.}\label{fig4}
\end{figure}

Since no statistically significant excess over the SM prediction is seen in the current IceCube data, we use this information to put an upper bound on the $|\lambda'_{ijk}|$ couplings. To this effect, we perform a binned likelihood analysis with the Poisson likelihood function
\begin{align}
L \ = \ \prod_{{\rm bins}~i} \frac{e^{-\lambda_i}\lambda_i^{n_i}}{n_i!},
\end{align}
where the observed count $n_i$ in each bin $i$ is compared to the theory prediction 
$\lambda_i$, including the RPV contribution induced by $\lambda'_{ijk}$. We then construct a test statistic 
\begin{align}
-2\Delta \ln L \ = \ -2(\ln L-\ln L_{\rm max}),
\end{align}
from which a $1\sigma$  upper limit on $|\lambda'_{ijk}|$ corresponding to the value of $-2\Delta \ln L=1$ can be derived. Here $L_{\rm max}$ represents the likelihood value for $\lambda'_{ijk}=0$, i.e. with only the SM contribution to the interaction. Our results are shown in Fig.~\ref{fig4} (blue solid curves) for $|\lambda'_{11k}|$ and $|\lambda'_{12k}|$,~i.e. for electron-type neutrinos interacting with the 1st and 2nd generation quarks, respectively. 

For comparison, we also show the direct limits on $|\lambda'_{11k}|$ from direct searches in $e^\pm p$ collisions at HERA with $\sqrt s=319$ GeV~\cite{Collaboration:2010ez, South:2016cmx}, as shown by the magenta-shaded region in Fig.~\ref{fig4} (left). Squark masses below 100 GeV or so are disfavored from direct searches for RPV SUSY at LEP~\cite{Abbiendi:2001aj, Heister:2002jc, Abbiendi:2003rn}, Tevatron~\cite{Abbott:1999nh, Abe:1998gu} and LHC~\cite{atlas-susy, cms-susy}, and therefore, are not considered here.\footnote{In the absence of the possibility of a resonant production (as e.g. in the sneutrino case) in $e^+e^-$ and hadron-hadron collisions, it is difficult to cast most of the collider limits onto the $m_{\tilde{d}}-|\lambda'_{ijk}|$ plane in a model-independent manner, and therefore, we do not attempt to show them in Fig.~\ref{fig4}.} In addition, the recent search for scalar leptoquarks at the 13 TeV LHC~\cite{Aaboud:2019jcc} is relevant for our RPV scenario, since $\lambda'_{ijk}$-couplings also give rise to the same $e_i e_ijj$ final states via pair-production of down-type squarks (from gluon fusion), followed by $\tilde{d}_{kR}\to e_{iL}u_{jL}$ which has a branching ratio of 0.5. The corresponding 95\% confidence level (CL) ATLAS limit of $\sim 1.2$ TeV on the first-generation scalar leptoquark mass can be directly translated into a lower bound on the down-type squark mass, as shown by the vertical dot-dashed line in Fig.~\ref{fig4}.  There also exist indirect constraints on $|\lambda'_{11k}|$ from lepton universality in pion decay, measured by the ratio $R_\pi=\frac{{\rm BR}(\pi^-\to e^-\bar{\nu}_e)}{{\rm BR}(\pi^-\to \mu^-\bar{\nu}_\mu)}$, unitarity of the CKM element $V_{ud}$ and atomic parity violation~\cite{Kao:2009fg}, the most stringent of which is shown in Fig.~\ref{fig4} (left) by the red dotted curve. Other low-energy constraints, such as neutrino mass~\cite{Bhattacharyya:1999tv}, electric dipole moment~\cite{Yamanaka:2014nba} and flavor-changing $B$-decays~\cite{Ghosh:2001mr, Dreiner:2013jta}, always involve the product of two independent RPV couplings, and hence, are not applicable in our case. For $k=1$, we have an additional constraint from $0\nu\beta\beta$, which however depends on the masses of other SUSY particles~\cite{Rodejohann:2011mu}, unlike the other limits discussed above, which are independent of the rest of the SUSY spectrum, as long as the 2-body RPV decay modes of the squark are dominant. Just for the sake of comparison, we translate the $0\nu\beta\beta$ half-life limit on $^{76}$Ge from  GERDA phase-II~\cite{Agostini:2018tnm} onto our RPV parameter space in Fig.~\ref{fig4} (left) for two benchmark points with gluino and neutralino masses of 10 TeV (gray, dashed) and 10 PeV (gray, dotted), while keeping all sfermion masses equal to $m_{\tilde{d}}$. In the former case, the $0\nu\beta\beta$ limit is the most stringent one, whereas for either heavier gaugino masses or $k\neq 1$, the pion decay constraint is stronger than the limit obtained from IceCube in the entire mass range considered. Similarly, as shown in Fig.~\ref{fig4} (right), for $|\lambda'_{12k}|$, the indirect constraints from lepton universality in neutral and charged $D$-meson decays, measured by the ratio $R_D=\frac{{\rm BR}(D\to Ke\nu_e)}{{\rm BR}(D\to K\mu\nu_\mu)}$,  
are stronger than the IceCube limit derived here. This rules out the possibility of any $|\lambda'_{1jk}|$-induced observable excess in the 8-year IceCube data.

We note that the IceCube constraints are mostly limited by statistics, which is  expected to improve 
significantly with more exposure time. To illustrate this point, we just 
scale the current 8-year dataset by a factor of 4 (roughly corresponding to 
30 years of actual data taking) in all the bins analyzed here and derive 
projected limits on $|\lambda'_{1jk}|$ (with $j=1,2$), following the same 
likelihood procedure described above. This conservative estimate of the future limits is shown in Fig.~\ref{fig4} by 
the blue dashed curves. We find that the limit on $|\lambda'_{1jk}|$ 
can be improved roughly by up to a factor of 3, and it might even surpass the current best limit in the sub-TeV squark mass range for $j=2$, although the indirect limit from lepton universality could improve by an order of magnitude at Belle II~\cite{Kou:2018nap}. In practice, however, we may not have to 
wait for 30 years, since a number of unforeseen factors could improve the 
conservative projected IceCube limits shown here, e.g. the future data in all the bins may not scale proportionately to the current data and may turn out to be in better agreement
 with the SM prediction. Similarly, other large-volume detectors like KM3NeT~\cite{Adrian-Martinez:2016fdl} and 
IceCube-Gen2~\cite{Aartsen:2014njl} might go online soon,  thus significantly increasing the total statistics. 

We also note that a similar analysis could be performed for incident neutrinos of muon and tau flavors at IceCube, though for muon neutrinos, one has to carefully reassess the atmospheric background including the RPV effects and also take into account the large uncertainty in converting the deposited muon energy to actual incoming $\nu_\mu$ energy. In any case, we expect the corresponding limits on $|\lambda'_{ijk}|$ (with $i=2,3$ and $j=1,2$) to be weaker than the limits on $|\lambda'_{1jk}|$ shown in Fig.~\ref{fig4} simply due to the fact that the effective fiducial volume at the IceCube is the largest for $\nu_e$~\cite{Aartsen:2017mau}. Nevertheless, with more statistics, one could in principle consider the $|\lambda'_{ijk}|$ couplings for all neutrino flavors. Also, one could improve the analysis presented here by taking into account the showers and tracks individually. Since the $|\lambda'_{1jk}|$ couplings preferentially enhance {\it only} one type of events, viz. showers for $i=1,3$ and tracks for $i=2$, a binned track-to-shower ratio analysis is expected to improve the limits on the corresponding $|\lambda'_{ijk}|$. In fact, by examining the track-to-shower ratio in future data, one might be able to distinguish between different new physics contributions to the IceCube events, provided the source flavor composition of the neutrinos is known more accurately.

\section{Long-lived Charged Sparticles} \label{sec:5}
\begin{figure}[t]
\centering
\includegraphics[width=0.7\textwidth]{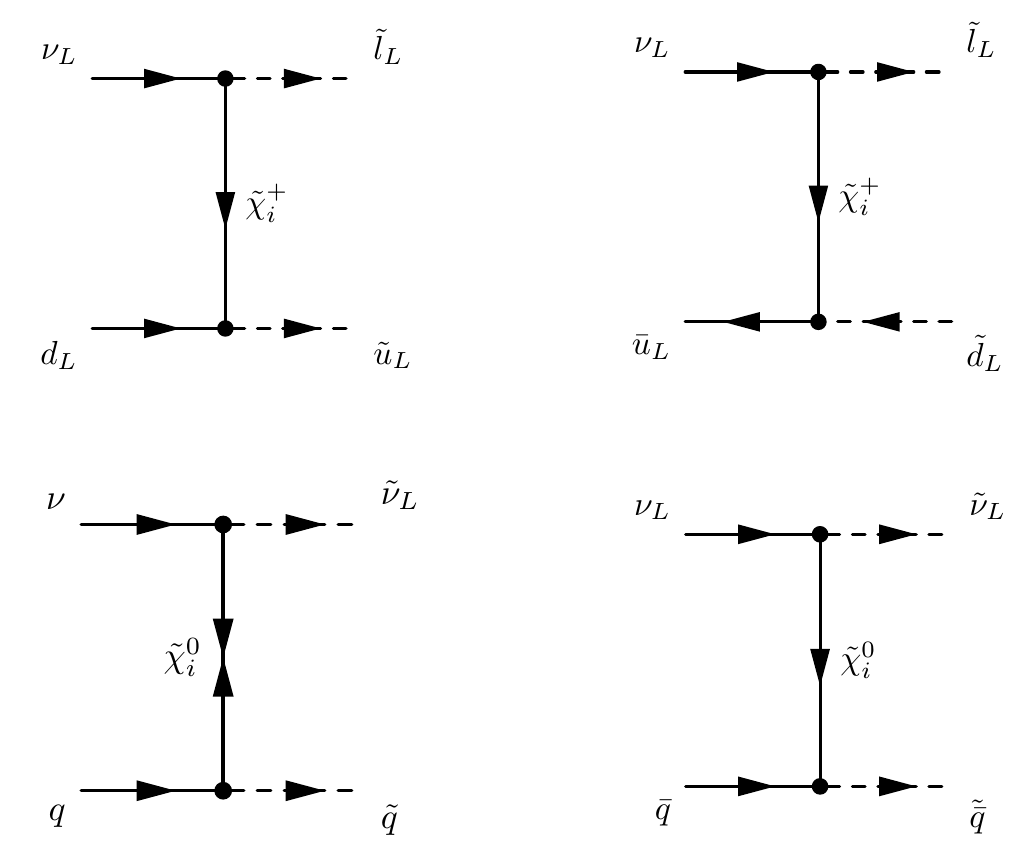}
\caption{Representative Feynman diagrams for NLSP production in neutrino-nucleon interaction in Earth without involving RPV interactions.} \label{fig:3}
\end{figure}

In SUSY theories where the LSP is the gravitino and the next-to-LSP (NLSP) is a long-lived charged particle (typically a slepton), e.g. in gauge mediation models~\cite{Giudice:1998bp}, the diffuse flux of UHE neutrinos interacting with the Earth could produce pairs of slepton NLSPs which could be detected in neutrino telescopes~\cite{Albuquerque:2003mi, Ahlers:2006pf, Albuquerque:2006am, Ando:2007ds, Canadas:2008ey,  Meade:2009mu}. The basic process involves the $t$-channel production of a slepton and a squark via gaugino exchange, see Fig.~\ref{fig:3}. Note that this does not require any RPV interactions.  The neutrino can interact either with an LH down-type quark, or with an RH up-type anti-quark, with the resulting partonic cross sections respectively given by 
\begin{align}
\frac{d\sigma_1}{dt} \ & = \ \frac{\pi \alpha}{2\sin^4\theta_w}\frac{m_{\widetilde{\chi}}^2}{s(t-m_{\widetilde{\chi}}^2)^2} \, , \\
\frac{d\sigma_2}{dt} \ & = \ \frac{\pi \alpha}{2\sin^4\theta_w}\frac{(tu-m^2_{\tilde{l}_L}m^2_{\tilde{q}})}{s^2(t-m_{\widetilde{\chi}}^2)^2} \, , 
\end{align}
where $s$, $t$ and $u$ are the usual Mandelstam variables, $\alpha\equiv e^2/4\pi$ is the fine structure constant, and $m_{\widetilde{\chi}}$ is the relevant gaugino (charged/neutral) mass. The heavier LH slepton and the squark promptly decay to the lighter RH slepton plus SM particles, thus giving rise to a pair of slepton NLSPs. The SUSY cross sections are considerably suppressed with respect to the SM neutrino-nucleon interactions, even when well above the threshold, unlike in the resonant case. This is because the SM cross section is dominated by very small values of $x$, whereas for the SUSY processes, we need $x>\frac{m^2_{\rm SUSY}}{2m_NE_\nu}$, where $m_{\rm SUSY}$ is the typical mass of the SUSY particle being produced.  However, if the NLSP slepton happens to be long-lived, as is the case in gauge-mediated SUSY breaking with stau NLSP, for instance, where the decay length in the lab frame is typically 
\begin{align}
c\tau \sim \left(\frac{\sqrt{F}}{10~{\rm PeV}}\right)^4 \left(\frac{100~{\rm GeV}}{m_{\tilde{\tau}_R}}\right)^5 10~{\rm km} \, ,
\end{align}
with $\sqrt{F}$ being the SUSY breaking scale, the long range of slepton compared to the background muon range of a few km compensates for the small SUSY production cross section. The high boost and large range of the NLSPs implies that its upward going tracks could in principle be detected in large ice or water Cherenkov detectors like IceCube or KM3NeT, provided one applies specialized cuts in the track separation to distinguish the signal events from the di-muon background. 

In analyzing this type of signal, it is very important to include the energy loss during the propagation of sleptons through the Earth. It turns out that for heavy charged particles, the bremsstrahlung and pair-production energy losses become less important, and the photo-nuclear energy loss is mass-suppressed for slepton masses much larger than the tau mass~\cite{Reno:2005si}. Thus, neutrino telescopes provide a complementary probe of long-lived charged slepton NLSP and intermediate-scale SUSY breaking, beyond the current collider limits of about 450 GeV~\cite{Khachatryan:2016sfv}. 

The long-lived stau hypothesis has gained some recent interest in view of two, anomalous upward-going ultra-high energy cosmic ray (UHECR) air shower events, with deposited shower energies of $0.6\pm 0.4$ EeV and $0.56^{+0.3}_{-0.2}$ EeV,  observed by the Antarctic Impulsive Transient Antenna (ANITA) balloon experiment~\cite{Gorham:2016zah, Gorham:2018ydl}. 
This is puzzling, because no SM particle is expected to survive passage through Earth a chord distance of $\sim 7000$ km (corresponding to the observed zenith angles of the two events) at EeV energies. In particular, the interpretation of these events as $\tau$-lepton decay-induced air showers at or near the ice surface arising from a diffuse UHE flux of cosmic $\nu_\tau$ is strongly disfavored due to their mean interaction length of only $\sim 300$ km. Including the effect of $\nu_\tau$ regeneration, the resulting survival probability over the chord length of the ANITA events with energy greater than $0.1 \; \text{EeV}$ is $<10^{-6}$, thus excluding the SM interpretation at $5.8\sigma$ CL~\cite{Fox:2018syq, Romero-Wolf:2018zxt}.  


In the gauge-mediated SUSY breaking models, the long-lived NLSP stau seems like an ideal candidate to explain the ANITA anomalous events~\cite{Connolly:2018ewv,Fox:2018syq}. Its cross section for nuclear interactions at $\sim$EeV energies is roughly 10 pb, three orders of magnitude less than the total neutrino cross section of 15 nb at 1 EeV, so that its mean free path through the Earth is several thousand km, while allowing for a reasonable branching ratios (BR $\lesssim 10^{-4}$) via UHECR neutrino-nucleon interactions~\cite{Albuquerque:2003mi, Albuquerque:2006am}. The stau NLSP then propagates across the required chord length of $\sim$ 7000 km and, for appropriate mass and lifetime, decays to a tau lepton plus  LSP prior to Earth emergence and subsequent extensive air shower. 

The main problem however is the energy loss of a heavy, charged particle like stau along its journey through the Earth due to ionization and radiative losses. At production, due to its high boost factor of ${\cal O}(10^7)$, the stau will have a lifetime that is heavily dilated in the lab frame, which however gradually decreases as it loses energy along its path. This was used as an advantage in~\cite{Connolly:2018ewv} to effectively reduce the boost factor, thus lowering the lab-frame decay time, causing the particle more likely to decay at the end of the chord length close to the surface of the Earth. At about 7000 km chord length (corresponding to the incident angle of the ANITA events), the rate of decay probability per distance $\frac{d\rm P_{\rm decay}}{dx}\sim 10^{-4}\ {\rm km}^{-1})$ for a TeV-scale stau~\cite{Connolly:2018ewv}. This could be translated into a decay probability around $0.01\%$ assuming that the decay happens within 40 km above the surface (corresponding to the flight height of ANITA) after exiting the Earth. However, this probability rate was calculated under the assumption that the incoming stau has an initial energy of $10^{22}\ \rm eV$, which is roughly two orders of magnitude above the Greisen-Zatsepin-Kuzmin (GZK) cut-off~\cite{Greisen:1966jv, Zatsepin:1966jv}. Lowering the initial energy of stau would bring down the decay probability even more. Thus, there is a competing effect between the decay probability rate and the energy loss rate, which makes it difficult, if not impossible, for the long-lived stau scenario to explain the anomalous ANITA events. In any case, this is yet another interesting possibility for probing SUSY at neutrino telescopes that should be further looked into.

\section{Long-lived Neutral Sparticles} \label{sec:6}
\begin{figure}[t]
\centering
\includegraphics[width=0.7\textwidth]{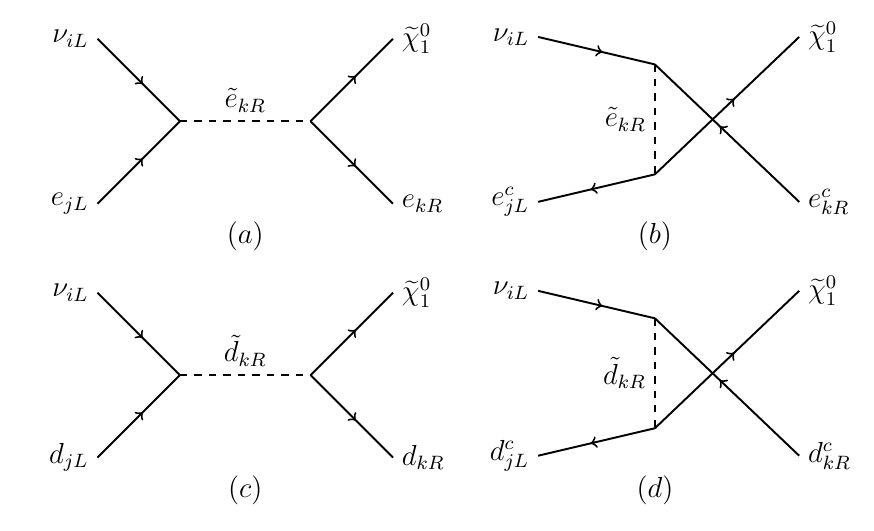}
\caption{Representative Feynman diagrams for the neutrino-electron (top) and neutrino-nucleon (bottom) interactions via RH-sfermion mediation in the RPV-SUSY scenario to produce a long-lived bino ($\widetilde{\chi}_1^0$) NLSP.} \label{fig:bino}
\end{figure}

If the NLSP is electrically neutral (e.g. neutralino or sneutrino) and long-lived, then it  has some advantages over the charged NLSP, as far as its detection at neutrino telescopes goes. In particular, a light bino in RPV SUSY turns out to be a natural candidate for explaining the anomalous ANITA events~\cite{Collins:2018jpg}. The $LQD$ (LLE)-type RPV couplings allow the on-shell, resonant production of a TeV-scale squark (slepton) from neutrino-nucleon (electron) scattering inside Earth, see Fig.~\ref{fig:bino}, thereby naturally enhancing the signal cross section. The squark/slepton decay inside the Earth can produce a pure bino which interacts with the SM fermions only via the $U(1)_Y$ gauge interactions and heavier SUSY particles, and therefore, can easily travel through thousands of km inside the Earth without significant energy loss, unlike tau, stau or any other electrically-charged particle. For a suitable, yet realistic sparticle mass spectrum and couplings consistent with all existing low and high-energy constraints, we find some parameter space where the bino is sufficiently long-lived (with proper lifetime of order of ns) and decays to SM fermions at or near the exit-surface of Earth to induce the air shower observed by ANITA~\cite{Collins:2018jpg}. For $LLE$-type RPV couplings [cf. Eq.~\eqref{eq:LLE}], the bino decays to a $\tau$-lepton (if kinematically allowed) and electron, either of which could induce the air shower seen by ANITA, whereas for $LQD$-type couplings [cf. Eq.~\eqref{laglamp}], the bino directly decays to quarks and neutrino, which mimic the SM $\tau$-decay. In the latter case, there exist some parameter space for which no throughgoing track events are predicted at IceCube. This might explain why IceCube has not seen either of the EeV events seen by ANITA, despite of its higher exposure time. 

Considering the $LLE$-type interactions first [cf.~Eq.~\eqref{eq:LLE}], the resonantly-produced slepton can decay back to an electron and neutrino through the same RPV interaction in Eq.~\eqref{eq:LLE} or to the corresponding lepton and neutralino through gauge interactions. The slepton might in principle be from any generation, though here we will make the assumption that the slepton is a stau, and also assume the lightest neutralino ($\chi_1^0$) to be much lighter than the stau, so that the decay $\widetilde{\tau}\to \tau \widetilde{\chi}_1^0$ is kinematically allowed. All other sparticles are assumed to be heavier than the stau and do not play any role, except for the gravitino ($\widetilde{G}$), which could be the LSP and plays the role of DM in this scenario. Gravitino LSP and bino NLSP can be realized, e.g. in natural gauge mediation without gaugino unification~\cite{Barnard:2012au}.

The cross-section for $\nu e\to \widetilde{\tau}\to \tau\widetilde{\chi}_1^0$ is given by~\cite{Carena:1998gd} 
\begin{align}
\sigma_{LLE} \ = \ \frac{8\pi s}{m^2_{\widetilde{\tau}}} \frac{\Gamma(\widetilde{\tau}\to e\nu)\ \Gamma(\widetilde{\tau}\to \tau\widetilde{\chi}_1^0)}{(s-m^2_{\widetilde{\tau}})^2+m_{\widetilde{\tau}}^2\Gamma_{\widetilde{\tau}}^2}\left(\frac{s-m^2_{\widetilde{\chi}_1^0}}{m^2_{\widetilde{\tau}}-m^2_{\widetilde{\chi}_1^0}}\right)^2 \, ,
\end{align}
which can be approximated by a Breit-Wigner formula close to the $s$-pole with $s\to m^2_{\widetilde{\tau}}$: 
\begin{align}
\sigma_{LLE}& \ \simeq \ \frac{8\pi}{m_{\widetilde{\tau}}^2}{\rm Br}(\widetilde{\tau}\to\nu e)\ {\rm Br}(\widetilde{\tau}\to\widetilde{\chi}_1^0\tau) 
 \ = \ \frac{8\pi}{m_{\widetilde{\tau}}^2}\frac{\lvert \lambda_{ijk}\rvert^2 g'^2}{(\lvert\lambda_{ijk}\rvert^2+g'^2)^2} \, ,\label{RPVXS}
\end{align} 
where $g'\equiv e/\cos\theta_w$ is the $U(1)_Y$ gauge coupling, and $j=1$, $k=3$ or vice versa, depending on whether it is the RH or LH-slepton resonance, respectively. The index $i$ for the incoming neutrino is free and we will assume a democratic flux ratio $1:1:1$ for $\nu_e:\nu_\mu:\nu_\tau$ and similarly for antineutrinos, as expected for a typical astrophysical neutrino flux with $1:2:0$ flavor composition at the source~\cite{Learned:1994wg}. 

We assume the bino is long-lived enough to survive its passage through Earth, before decaying close to or at the surface of exit. It can decay back to the $\tau$-lepton and an off-shell stau, leading to a 3-body final state: $\widetilde{\chi}_1^0\to \tau^+ \widetilde{\tau}^{*-} \to \tau^+ e^- \bar{\nu}$. In principle, the upgoing shower may be initiated either directly by the electron, or by the subsequent decay of the $\tau$. In the limit $m_{\widetilde{\chi}_1^0}\ll m_{\widetilde{\tau}}$, the 3-body decay rate can be estimated as 
\begin{equation}\label{decaywidth}
\Gamma(\widetilde{\chi}_1^0\to\tau^- e^+ \bar{\nu}) \ \simeq \ \frac{g'^2|\lambda_{ijk}|^2}{512\pi^3}\frac{m_{\widetilde{\chi}_1^0}^5}{m_{\widetilde{\tau}}^4} \, .
\end{equation}
The chord length of $\sim 7000$ km required for the two ANITA events translates into the mean lifetime of bino in the lab frame as $\tau_{\widetilde{\chi}_1^0}^{\rm lab}\sim 0.022$ s. From the decay kinematics of the event, we estimate that the incoming neutrino energy should be roughly four times the detected energy at ANITA. Given that the two ANITA events had an average energy of 0.5 EeV, the initial neutrino energy should be $E_\nu\sim$ 2 EeV. Then the resonance condition fixes the stau mass: $m_{\widetilde{\tau}}=\sqrt{2E_\nu m_e}\simeq 2$ TeV. Substituting this in Eq.~\eqref{decaywidth}, we find that for a typical value of $|\lambda|\sim 0.1$, as allowed by current experimental constraints~\cite{Kao:2009fg}, one needs a light bino of mass $m_{\widetilde{\chi}_1^0}\sim 8$ GeV. 

\begin{figure*}[t!]
	\centering
	\includegraphics[width=0.49\textwidth]{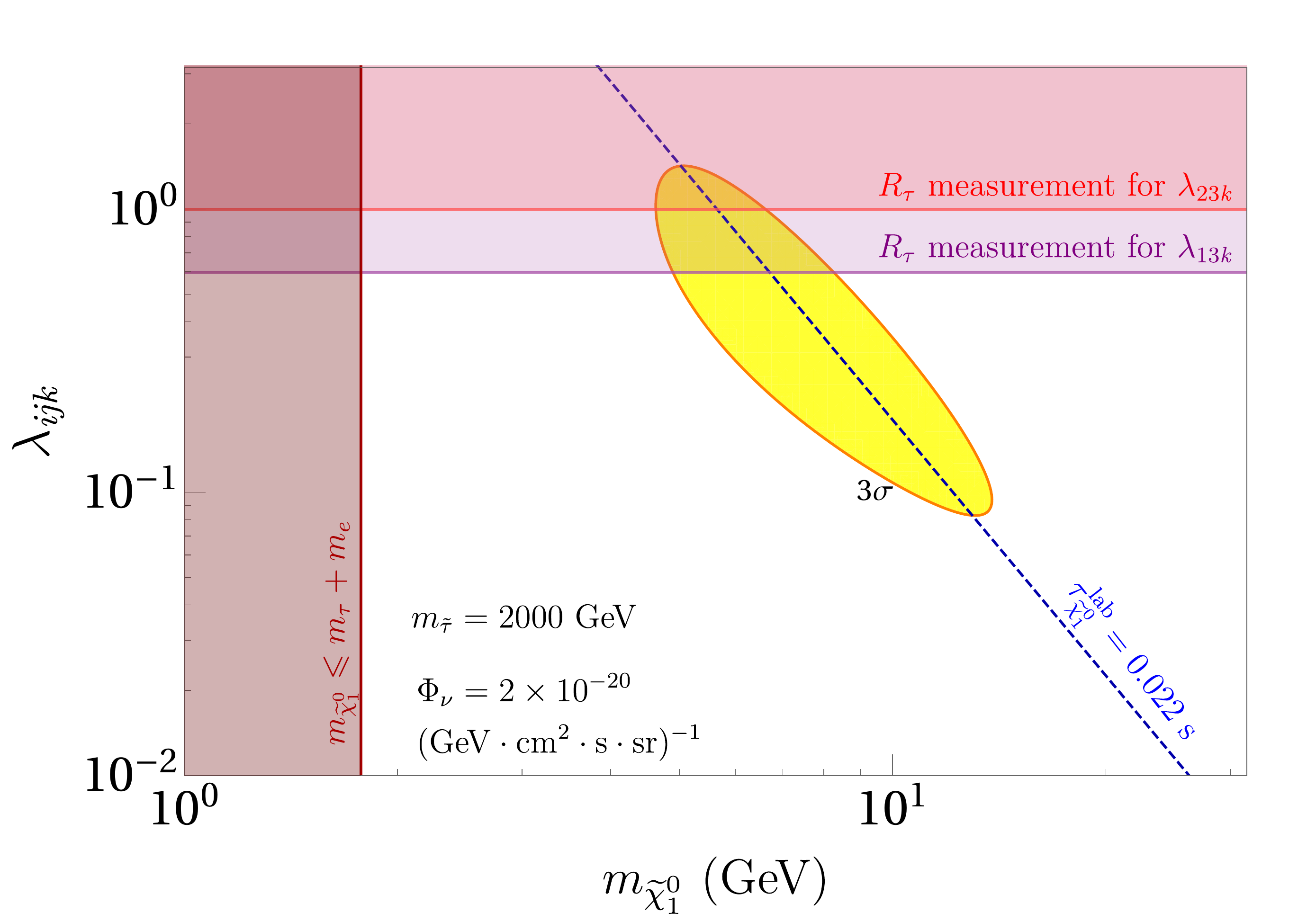}
\includegraphics[width=0.49\textwidth]{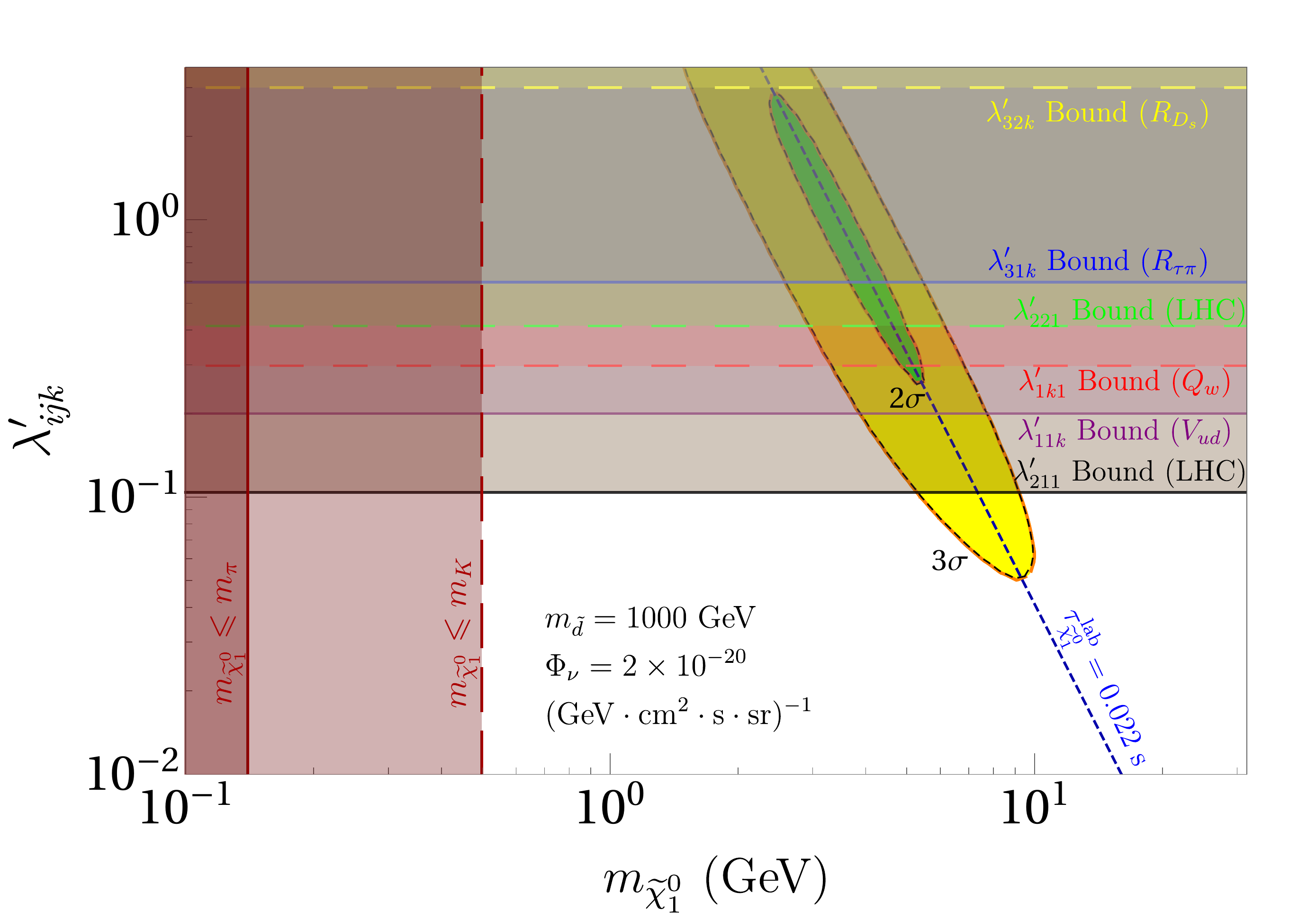}
	\caption{The 3$\sigma$ preferred region (yellow shaded) explaining the ANITA anomalous events in our RPV-SUSY framework. The left panel is for the $LLE$ case with a stau mass of $m_{\widetilde{\tau}}=2$ TeV, and the right panel is for the $LQD$ case with a down-squark mass of $m_{\widetilde{d}}=1$ TeV. The horizontal shaded areas are the excluded regions for single RPV couplings from low-energy experiments~\cite{Kao:2009fg}. The vertical shaded regions are the kinematically forbidden regions for the bino decay considered here. Taken from~\cite{Collins:2018jpg}.}
	\label{fig:lambdaMconstriant}
\end{figure*}

A more accurate calculation, taking into account all statistical interaction/decay probabilities for the neutrino, bino and tau, yields the $3\sigma$ allowed range in the $(m_{\widetilde{\chi}_1^0}, |\lambda|)$ plane that explains ANITA,  as shown by the yellow shaded region in Fig.~\ref{fig:lambdaMconstriant} (left panel)~\cite{Collins:2018jpg}.  Here we have assumed an anisotropic UHE neutrino source with flux $\Phi_\nu\sim 2\times10^{-20}\  (\rm GeV\cdot cm^2\cdot s\cdot sr)^{-1}$ and the mass of the RPV-SUSY mediator stau at $\rm m_{\widetilde{\tau}}=2$ TeV. The dashed blue line shows the relation between the parameters once we set the bino decay length to be exactly the maximum distance traveled inside Earth, which corresponds to a mean lab-frame lifetime of $\tau_{\widetilde{\chi}_1^0}^{\rm lab}\sim 0.022 
$ s. The horizontal purple and red shaded regions are excluded from the $R_\tau$ measurements~\cite{Kao:2009fg}. The vertical shaded region is the kinematically forbidden region for the bino to decay into a $\tau$-lepton. 
The stau mass is roughly fixed to lie in the $1\text{--}2 \; \text{TeV}$ region by the requirement that $m_{\tilde{\tau}} = \sqrt{s} = \sqrt{2 m_e E_\nu}$, and that $E_\nu$ should be a few times larger than the observed cosmic ray energy of $0.2\text{--}1 \; \text{EeV}$. Such a particle may be observed in current or future collider experiments. The current LHC lower limits on the stau mass in the RPV-SUSY scenario is about 500 GeV, derived from multilepton searches~\cite{Aad:2014iza}. 

Now consider the $LQD$-type interactions [cf.~Eq.~\eqref{laglamp}]]. For simplicity, we only take the first-generation squark in the intermediate state. As for the initial state quarks, both $d$ and $s$ quark contributions turn out to be comparable. However, due to stringent constraints on the product $\lambda'_{i1k}\lambda'_{j2k}\lesssim 5\times 10^{-5}$ from $K$-meson  decays~\cite{Barbier:2004ez}, we will consider either the down-quark or the strange-quark in the initial state separately, but not both simultaneously. In particular, we will only consider the $\lambda'_{ijk}$ couplings with $j=1,2$ and $k=1$ for RH down-squark. After being resonantly produced, the bino will have a 3-body decay via off-shell down-type squark: $\widetilde{\chi}_1^0\to d \bar{d} \nu$ and $\widetilde{\chi}_1^0 \to u \bar{d} e$. 
In this case, the final-state quarks from the 3-body bino decay  hadronize to either pions or kaons, mimicking the hadronic shower induced by the $\tau$. 

The total differential cross section for the neutrino-nucleon and antineutrino-nucleon interactions are respectively given by [cf.~Eq.~\eqref{cross}]~\cite{Collins:2018jpg} 
 \begin{align}
\frac{d\sigma^\nu_{LQD}}{dxdy} \ = \ &\frac{m_NE_{\nu}}{16\pi}\frac{|\lambda'_{ijk}|^2g'^2}{18}\left[ \frac{4xf_{d}(x,Q^2)}{\left(xs-m_{\tilde{d}_{R}}^2\right)^2+m^2_{\tilde{d}_{R}}\Gamma^2_{\tilde{d}_{R}}}
+  \frac{xf_{\bar{d}}(x,Q^2)}{\left(xs-m_{\tilde{d}_{L}}^2\right)^2+m^2_{\tilde{d}_{L}}\Gamma^2_{\tilde{d}_{L}}}\right] \, , \label{lqd1} \\
\frac{d\sigma^{\bar{\nu}}_{LQD}}{dxdy}\ = \ &\frac{m_NE_{\nu}}{16\pi}\frac{|\lambda'_{ijk}|^2g'^2}{18}\left[ \frac{xf_{d}(x,Q^2)}{\left(xs-m_{\tilde{d}_{L}}^2\right)^2+m^2_{\tilde{d}_{L}}\Gamma^2_{\tilde{d}_{L}}} 
+ \frac{4xf_{\bar{d}}(x,Q^2)}{\left(xs-m_{\tilde{d}_{R}}^2\right)^2+m^2_{\tilde{d}_{R}}\Gamma^2_{\tilde{d}_{R}}}\right] . 
\label{lqd2}
\end{align} 
The Breit-Wigner resonance is regulated by the squark widths\footnote{The RH down-squark has two RPV decay modes: $\widetilde{d}_{kR}\to \nu_{iL}d_{jL}, \,  e_{iL}u_{jL}$, whereas the LH down-squark has only one: $\widetilde{d}_{kL}\to \nu_{iL}d_{jR}$. Similarly, for the $R$-parity conserving decays $\widetilde{d}\to d\widetilde{\chi}_1^0$, the RH squark coupling is twice that of the LH squark (due to different hypercharges).}
\begin{align}
\Gamma_{\tilde{d}_{kR}} \ = \ \frac{m_{\widetilde{d}_{kR}}}{8\pi}\left[\sum_{ij}|\lambda'_{ijk}|^2+\frac{2}{9}g'^2\right] \, , \\
\Gamma_{\tilde{d}_{kL}} \ = \ \frac{m_{\widetilde{d}_{kL}}}{16\pi}\left[\sum_{ij}|\lambda'_{ijk}|^2+\frac{1}{9}g'^2\right] \, .
\end{align}
Note that the resonance condition is satisfied for the incoming neutrino energy $E_\nu=m^2_{\widetilde{d}}/2m_Nx$, but due to the spread in the initial quark momentum fraction $x\in [0,1]$, the resonance peak is broadened and shifted above the threshold value $E_\nu^{\rm th}=m^2_{\widetilde{d}}/2m_N$, unlike the $LLE$ case where the resonance was much narrower. This is one of the reasons why the $LQD$ case allows for a larger parameter space than the $LLE$ case in explaining the ANITA events, as can be seen from Fig.\ref{fig:lambdaMconstriant} (right panel). 

Our results for the $3\sigma$ preferred region in the $LQD$ case are shown in the right panel of Fig.~\ref{fig:lambdaMconstriant} (yellow shaded region) for both $\nu-d$ (solid contours) and $\nu-s$ (overlapping dashed contours) initial states. The vertical shaded regions are the kinematically forbidden regions for the bino to decay into pions or kaons, corresponding to the $\lambda'_{i11}$ or $\lambda'_{i21}$ scenario, respectively. 
The horizontal shaded regions bounded by the purple and blue solid lines are excluded from the $V_{ud}$ and $R_{\tau\pi}$ measurements, respectively~\cite{Kao:2009fg}, which constrain the $\nu-d$ scenario. Similarly, the shaded regions bounded by the red and yellow dashed lines are excluded from the $Q_w$ and $R_{D_s}$ measurements, respectively~\cite{Kao:2009fg}, which constrain the $\nu-s$ scenario. Here we have chosen the RH down-squark mediator mass as $m_{\widetilde{d}}=1$ TeV. We do not include the LH squark contribution, because according to our estimates, the production cross section for a 1-TeV RH down-squark at the $\sqrt s=13$ TeV LHC is 6.2 fb, which is safely below the current upper limit of 13.5 fb~\cite{Aaboud:2017vwy}, whereas including the LH squark contribution increases the cross section to 15.5 fb. The black and green shaded regions in Fig.~\ref{fig:lambdaMconstriant} are the exclusion regions based on a recent update of the LHC constraints on the LQD couplings $\lambda'_{211}$ and on $\lambda'_{221}$, respectively~\cite{Bansal:2018dge}. 

Based on these bounds, we find that there is allowed parameter space at both $2\sigma$ and $3\sigma$ for the $\lambda'_{i21}$ scenario ($\nu-s$ initial state), whereas for the 
$\lambda'_{i11}$ scenario ($\nu-d$ initial state), there is only a small $3\sigma$ range with $\lambda'_{\rm i11}\sim 0.07-0.1$ and $m_{\widetilde{\chi}_1^0}\sim 7-10$ GeV  allowed. Increasing the squark mass moves the $2\sigma$ and $3\sigma$ contours to larger $\lambda'$ values, which are excluded by the $V_{ud}$ and $Q_w$ measurements~\cite{Kao:2009fg}. Thus, we predict that if our $LQD$-type RPV-SUSY interpretation of the ANITA events is correct, then a TeV-scale squark should be soon found at the LHC.    Another independent test of the allowed parameter space shown in Fig.~\ref{fig:lambdaMconstriant} might come soon from the Belle II upgrade~\cite{Kou:2018nap}, which could significantly improve the $R_\tau$ measurements.

\section{Conclusion} \label{sec:7}
RPV SUSY is a well-motivated candidate for TeV-scale new physics beyond the SM, while being consistent with the null results at the LHC so far. Therefore, it is important to test this hypothesis at different energy scales available to us. We have reviewed the prospects of probing RPV SUSY at neutrino telescopes using the highest-energy neutrinos given to us by Nature. We discussed three different aspects of the SUSY models that can be tested in this way: (i) If there exists a TeV-scale squark/slepton, it could be resonantly produced via neutrino interactions with Earth matter, thus leading to a potentially observable excess in the UHE neutrino events at IceCube; (ii) If there exists a long-lived charged NLSP, it could lead to distinct upward-going tracks or air showers that could be detected by IceCube and balloon-borne experiments like ANITA, respectively; and (iii)  If there exists a long-lived neutral NLSP, it could also lead to extensive air shower events at ANITA-like experiments, while mostly evading the IceCube-detection (until we accumulate more data).  Thus, neutrino telescopes complement the RPV SUSY searches at the energy and intensity frontiers. 
It would be remarkable if weak-scale SUSY was discovered in such an unexpected way!

\section*{Acknowledgments}
BD thanks Chien-Yi Chen, Jack Collins, Dilip Ghosh, Werner Rodejohann, Amarjit Soni, and Yicong Sui for collaboration on projects related to the topics discussed here. He also thanks  Zackaria Chacko and Lucien Heurtier for useful discussions. Special thanks are due to Yicong Sui for help with Figure 2. This work was supported by the US Department of Energy under Grant No. DE-SC0017987 and also by the Neutrino Theory Network Program under Grant No. DE-AC02-07CH11359. 
\bibliographystyle{ws-rv-van}
\bibliography{dev_susy}

\begin{thebibliography}{150}
\providecommand{\natexlab}[1]{#1}
\providecommand{\url}[1]{\texttt{#1}}
\expandafter\ifx\csname urlstyle\endcsname\relax
  \providecommand{\doi}[1]{doi: #1}\else
  \providecommand{\doi}{doi: \begingroup \urlstyle{rm}\Url}\fi

\bibitem{Tanabashi:2018oca}
M.~Tanabashi et~al., {Review of Particle Physics}, \emph{Phys. Rev.} {\bf
  D98}\penalty0 (3), \penalty0 030001  (2018).
\newblock \doi{10.1103/PhysRevD.98.030001}.

\bibitem{Haber:1984rc}
H.~E. Haber and G.~L. Kane, {The Search for Supersymmetry: Probing Physics
  Beyond the Standard Model}, \emph{Phys. Rept.} {\bf 117}, \penalty0 75--263
  (1985).
\newblock \doi{10.1016/0370-1573(85)90051-1}.

\bibitem{Baer:2006rs}
H.~Baer and X.~Tata, \emph{{Weak scale supersymmetry: From superfields to
  scattering events}}. Cambridge University Press  (2006).
\newblock URL \url{http://www.cambridge.org/9780521290319}.

\bibitem{atlas-susy}
\url{https://twiki.cern.ch/twiki/bin/view/AtlasPublic/SupersymmetryPublicResults}.

\bibitem{cms-susy}
\url{https://twiki.cern.ch/twiki/bin/view/CMSPublic/PhysicsResultsSUS}.

\bibitem{Papucci:2011wy}
M.~Papucci, J.~T. Ruderman, and A.~Weiler, {Natural SUSY Endures}, \emph{JHEP}.
  {\bf 09}, \penalty0 035  (2012).
\newblock \doi{10.1007/JHEP09(2012)035}.

\bibitem{Hall:2011aa}
L.~J. Hall, D.~Pinner, and J.~T. Ruderman, {A Natural SUSY Higgs Near 126 GeV},
  \emph{JHEP}. {\bf 04}, \penalty0 131  (2012).
\newblock \doi{10.1007/JHEP04(2012)131}.

\bibitem{Brust:2011tb}
C.~Brust, A.~Katz, S.~Lawrence, and R.~Sundrum, {SUSY, the Third Generation and
  the LHC}, \emph{JHEP}. {\bf 03}, \penalty0 103  (2012).
\newblock \doi{10.1007/JHEP03(2012)103}.

\bibitem{Buckley:2016kvr}
M.~R. Buckley, D.~Feld, S.~Macaluso, A.~Monteux, and D.~Shih, {Cornering
  Natural SUSY at LHC Run II and Beyond}, \emph{JHEP}. {\bf 08}, \penalty0 115
  (2017).
\newblock \doi{10.1007/JHEP08(2017)115}.

\bibitem{Ahmed:2017jxl}
W.~Ahmed, X.-J. Bi, T.~Li, J.~S. Niu, S.~Raza, Q.-F. Xiang, and P.-F. Yin,
  {Status of natural supersymmetry from the generalized minimal supergravity in
  light of the current LHC run-2 and LUX data}, \emph{Phys. Rev.} {\bf
  D98}\penalty0 (1), \penalty0 015040  (2018).
\newblock \doi{10.1103/PhysRevD.98.015040}.

\bibitem{Barbier:2004ez}
R.~Barbier et~al., {R-parity violating supersymmetry}, \emph{Phys. Rept.} {\bf
  420}, \penalty0 1--202  (2005).
\newblock \doi{10.1016/j.physrep.2005.08.006}.

\bibitem{Carpenter:2006hs}
L.~M. Carpenter, D.~E. Kaplan, and E.-J. Rhee, {Reduced fine-tuning in
  supersymmetry with R-parity violation}, \emph{Phys. Rev. Lett.} {\bf 99},
  \penalty0 211801  (2007).
\newblock \doi{10.1103/PhysRevLett.99.211801}.

\bibitem{Dreiner:2012np}
H.~K. Dreiner and T.~Stefaniak, {Bounds on R-parity Violation from Resonant
  Slepton Production at the LHC}, \emph{Phys. Rev.} {\bf D86}, \penalty0 055010
   (2012).
\newblock \doi{10.1103/PhysRevD.86.055010}.

\bibitem{Allanach:2012vj}
B.~C. Allanach and B.~Gripaios, {Hide and Seek With Natural Supersymmetry at
  the LHC}, \emph{JHEP}. {\bf 05}, \penalty0 062  (2012).
\newblock \doi{10.1007/JHEP05(2012)062}.

\bibitem{Dreiner:2012wm}
H.~K. Dreiner, F.~Staub, A.~Vicente, and W.~Porod, {General MSSM signatures at
  the LHC with and without R-parity}, \emph{Phys. Rev.} {\bf D86}, \penalty0
  035021  (2012).
\newblock \doi{10.1103/PhysRevD.86.035021}.

\bibitem{Asano:2012gj}
M.~Asano, K.~Rolbiecki, and K.~Sakurai, {Can R-parity violation hide vanilla
  supersymmetry at the LHC?}, \emph{JHEP}. {\bf 01}, \penalty0 128  (2013).
\newblock \doi{10.1007/JHEP01(2013)128}.

\bibitem{Graham:2014vya}
P.~W. Graham, S.~Rajendran, and P.~Saraswat, {Supersymmetric crevices: Missing
  signatures of R -parity violation at the LHC}, \emph{Phys. Rev.} {\bf
  D90}\penalty0 (7), \penalty0 075005  (2014).
\newblock \doi{10.1103/PhysRevD.90.075005}.

\bibitem{Monteux:2016gag}
A.~Monteux, {New signatures and limits on R-parity violation from resonant
  squark production}, \emph{JHEP}. {\bf 03}, \penalty0 216  (2016).
\newblock \doi{10.1007/JHEP03(2016)216}.

\bibitem{Dercks:2017lfq}
D.~Dercks, H.~Dreiner, M.~E. Krauss, T.~Opferkuch, and A.~Reinert, {R-Parity
  Violation at the LHC}, \emph{Eur. Phys. J.} {\bf C77}\penalty0 (12),
  \penalty0 856  (2017).
\newblock \doi{10.1140/epjc/s10052-017-5414-4}.

\bibitem{Guo:2018hbv}
J.~Guo, J.~Li, T.~Li, F.~Xu, and W.~Zhang, {Deep learning for $R$-parity
  violating supersymmetry searches at the LHC}, \emph{Phys. Rev.} {\bf
  D98}\penalty0 (7), \penalty0 076017  (2018).
\newblock \doi{10.1103/PhysRevD.98.076017}.

\bibitem{Bansal:2018dge}
S.~Bansal, A.~Delgado, C.~Kolda, and M.~Quiros, {Limits on R-parity-violating
  couplings from Drell-Yan processes at the LHC}, \emph{Phys. Rev.} {\bf
  D99}\penalty0 (9), \penalty0 093008  (2019).
\newblock \doi{10.1103/PhysRevD.99.093008}.

\bibitem{Hall:1983id}
L.~J. Hall and M.~Suzuki, {Explicit R-Parity Breaking in Supersymmetric
  Models}, \emph{Nucl. Phys.} {\bf B231}, \penalty0 419--444  (1984).
\newblock \doi{10.1016/0550-3213(84)90513-3}.

\bibitem{Ellis:1984gi}
J.~R. Ellis, G.~Gelmini, C.~Jarlskog, G.~G. Ross, and J.~W.~F. Valle,
  {Phenomenology of Supersymmetry with Broken R-Parity}, \emph{Phys. Lett.}
  {\bf 150B}, \penalty0 142--148  (1985).
\newblock \doi{10.1016/0370-2693(85)90157-1}.

\bibitem{Dawson:1985vr}
S.~Dawson, {R-Parity Breaking in Supersymmetric Theories}, \emph{Nucl. Phys.}
  {\bf B261}, \penalty0 297--318  (1985).
\newblock \doi{10.1016/0550-3213(85)90577-2}.

\bibitem{Joshipura:1994ib}
A.~S. Joshipura and M.~Nowakowski, {'Just so' oscillations in supersymmetric
  standard model}, \emph{Phys. Rev.} {\bf D51}, \penalty0 2421--2427  (1995).
\newblock \doi{10.1103/PhysRevD.51.2421}.

\bibitem{Hempfling:1995wj}
R.~Hempfling, {Neutrino masses and mixing angles in SUSY GUT theories with
  explicit R-parity breaking}, \emph{Nucl. Phys.} {\bf B478}, \penalty0 3--30
  (1996).
\newblock \doi{10.1016/0550-3213(96)00412-9}.

\bibitem{Roy:1996bua}
S.~Roy and B.~Mukhopadhyaya, {Some implications of a supersymmetric model with
  R-parity breaking bilinear interactions}, \emph{Phys. Rev.} {\bf D55},
  \penalty0 7020--7029  (1997).
\newblock \doi{10.1103/PhysRevD.55.7020}.

\bibitem{Mukhopadhyaya:1998xj}
B.~Mukhopadhyaya, S.~Roy, and F.~Vissani, {Correlation between neutrino
  oscillations and collider signals of supersymmetry in an R-parity violating
  model}, \emph{Phys. Lett.} {\bf B443}, \penalty0 191--195  (1998).
\newblock \doi{10.1016/S0370-2693(98)01288-X}.

\bibitem{Bhattacharyya:1999tv}
G.~Bhattacharyya, H.~V. Klapdor-Kleingrothaus, and H.~Pas, {Neutrino mass and
  magnetic moment in supersymmetry without R parity in the light of recent
  data}, \emph{Phys. Lett.} {\bf B463}, \penalty0 77--82  (1999).
\newblock \doi{10.1016/S0370-2693(99)00947-8}.

\bibitem{Davidson:2000ne}
S.~Davidson and M.~Losada, {Basis independent neutrino masses in the R(p)
  violating MSSM}, \emph{Phys. Rev.} {\bf D65}, \penalty0 075025  (2002).
\newblock \doi{10.1103/PhysRevD.65.075025}.

\bibitem{Diaz:2003as}
M.~A. Diaz, M.~Hirsch, W.~Porod, J.~C. Romao, and J.~W.~F. Valle, {Solar
  neutrino masses and mixing from bilinear R parity broken supersymmetry:
  Analytical versus numerical results}, \emph{Phys. Rev.} {\bf D68}, \penalty0
  013009  (2003).
\newblock \doi{10.1103/PhysRevD.71.059904, 10.1103/PhysRevD.68.013009}.
\newblock [Erratum: Phys. Rev.D71,059904(2005)].

\bibitem{Dreiner:2011ft}
H.~K. Dreiner, M.~Hanussek, J.-S. Kim, and C.~H. Kom, {Neutrino masses and
  mixings in the baryon triality constrained minimal supersymmetric standard
  model}, \emph{Phys. Rev.} {\bf D84}, \penalty0 113005  (2011).
\newblock \doi{10.1103/PhysRevD.84.113005}.

\bibitem{Peinado:2012tp}
E.~Peinado and A.~Vicente, {Neutrino Masses from R-Parity Violation with a
  $Z_3$ Symmetry}, \emph{Phys. Rev.} {\bf D86}, \penalty0 093024  (2012).
\newblock \doi{10.1103/PhysRevD.86.093024}.

\bibitem{Mohapatra:1986su}
R.~N. Mohapatra, {New Contributions to Neutrinoless Double beta Decay in
  Supersymmetric Theories}, \emph{Phys. Rev.} {\bf D34}, \penalty0 3457--3461
  (1986).
\newblock \doi{10.1103/PhysRevD.34.3457}.
\newblock [,778(1986)].

\bibitem{Vergados:1986td}
J.~D. Vergados, {Neutrinoless Double Beta Decay Without Majorana Neutrinos in
  Supersymmetric Theories}, \emph{Phys. Lett.} {\bf B184}, \penalty0 55--62
  (1987).
\newblock \doi{10.1016/0370-2693(87)90487-4}.

\bibitem{Hirsch:1995zi}
M.~Hirsch, H.~V. Klapdor-Kleingrothaus, and S.~G. Kovalenko, {New constraints
  on R-parity broken supersymmetry from neutrinoless double beta decay},
  \emph{Phys. Rev. Lett.} {\bf 75}, \penalty0 17--20  (1995).
\newblock \doi{10.1103/PhysRevLett.75.17}.

\bibitem{Babu:1995vh}
K.~S. Babu and R.~N. Mohapatra, {New vector - scalar contributions to
  neutrinoless double beta decay and constraints on R-parity violation},
  \emph{Phys. Rev. Lett.} {\bf 75}, \penalty0 2276--2279  (1995).
\newblock \doi{10.1103/PhysRevLett.75.2276}.

\bibitem{Pas:1998nn}
H.~Pas, M.~Hirsch, and H.~V. Klapdor-Kleingrothaus, {Improved bounds on SUSY
  accompanied neutrinoless double beta decay}, \emph{Phys. Lett.} {\bf B459},
  \penalty0 450--454  (1999).
\newblock \doi{10.1016/S0370-2693(99)00711-X}.

\bibitem{Faessler:2007nz}
A.~Faessler, T.~Gutsche, S.~Kovalenko, and F.~Simkovic, {Pion dominance in RPV
  SUSY induced neutrinoless double beta decay}, \emph{Phys. Rev.} {\bf D77},
  \penalty0 113012  (2008).
\newblock \doi{10.1103/PhysRevD.77.113012}.

\bibitem{Allanach:2009xx}
B.~C. Allanach, C.~H. Kom, and H.~Pas, {LHC and B physics probes of
  neutrinoless double beta decay in supersymmetry without R-parity},
  \emph{JHEP}. {\bf 10}, \penalty0 026  (2009).
\newblock \doi{10.1088/1126-6708/2009/10/026}.

\bibitem{Rodejohann:2011mu}
W.~Rodejohann, {Neutrino-less Double Beta Decay and Particle Physics},
  \emph{Int. J. Mod. Phys.} {\bf E20}, \penalty0 1833--1930  (2011).
\newblock \doi{10.1142/S0218301311020186}.

\bibitem{Cline:1990bw}
J.~M. Cline and S.~Raby, {Gravitino induced baryogenesis: A Problem made a
  virtue}, \emph{Phys. Rev.} {\bf D43}, \penalty0 1781--1787  (1991).
\newblock \doi{10.1103/PhysRevD.43.1781}.

\bibitem{Masiero:1992bv}
A.~Masiero and A.~Riotto, {Cosmic Delta B from lepton violating interactions at
  the electroweak phase transition}, \emph{Phys. Lett.} {\bf B289}, \penalty0
  73--80  (1992).
\newblock \doi{10.1016/0370-2693(92)91364-F}.

\bibitem{Sarkar:1996sn}
U.~Sarkar and R.~Adhikari, {Baryogenesis through R-parity violation},
  \emph{Phys. Rev.} {\bf D55}, \penalty0 3836--3843  (1997).
\newblock \doi{10.1103/PhysRevD.55.3836}.

\bibitem{Dolgov:2006ay}
A.~D. Dolgov and F.~R. Urban, {Baryogenesis by R-parity violating top quark
  decays and neutron-antineutron oscillations}, \emph{Nucl. Phys.} {\bf B752},
  \penalty0 297--315  (2006).
\newblock \doi{10.1016/j.nuclphysb.2006.06.035}.

\bibitem{Kohri:2009ka}
K.~Kohri, A.~Mazumdar, and N.~Sahu, {Inflation, baryogenesis and gravitino dark
  matter at ultra low reheat temperatures}, \emph{Phys. Rev.} {\bf D80},
  \penalty0 103504  (2009).
\newblock \doi{10.1103/PhysRevD.80.103504}.

\bibitem{Cui:2012jh}
Y.~Cui and R.~Sundrum, {Baryogenesis for weakly interacting massive particles},
  \emph{Phys. Rev.} {\bf D87}\penalty0 (11), \penalty0 116013  (2013).
\newblock \doi{10.1103/PhysRevD.87.116013}.

\bibitem{Sorbello:2013xwa}
F.~Rompineve, {Weak Scale Baryogenesis in a Supersymmetric Scenario with
  R-parity violation}, \emph{JHEP}. {\bf 08}, \penalty0 014  (2014).
\newblock \doi{10.1007/JHEP08(2014)014}.

\bibitem{Arcadi:2015ffa}
G.~Arcadi, L.~Covi, and M.~Nardecchia, {Gravitino Dark Matter and low-scale
  Baryogenesis}, \emph{Phys. Rev.} {\bf D92}\penalty0 (11), \penalty0 115006
  (2015).
\newblock \doi{10.1103/PhysRevD.92.115006}.

\bibitem{Farina:2016ndq}
M.~Farina, A.~Monteux, and C.~S. Shin, {Twin mechanism for baryon and dark
  matter asymmetries}, \emph{Phys. Rev.} {\bf D94}\penalty0 (3), \penalty0
  035017  (2016).
\newblock \doi{10.1103/PhysRevD.94.035017}.

\bibitem{Pierce:2019ozl}
A.~Pierce and B.~Shakya, {Gaugino Portal Baryogenesis}  (2019).
\newblock 1901.05493.

\bibitem{Kim:2001se}
J.~E. Kim, B.~Kyae, and H.~M. Lee, {Effective supersymmetric theory and
  (g-2)(muon with R-parity violation}, \emph{Phys. Lett.} {\bf B520}, \penalty0
  298--306  (2001).
\newblock \doi{10.1016/S0370-2693(01)01134-0}.

\bibitem{Bhattacharyya:2009hb}
G.~Bhattacharyya, K.~B. Chatterjee, and S.~Nandi, {Correlated enhancements in
  D(s) ---> l nu, (g-2) of muon, and lepton flavor violating tau decays with
  two R-parity violating couplings}, \emph{Nucl. Phys.} {\bf B831}, \penalty0
  344--357  (2010).
\newblock \doi{10.1016/j.nuclphysb.2010.01.025}.

\bibitem{Hundi:2011si}
R.~S. Hundi, {Constraints from neutrino masses and muon (g-2) in the bilinear
  R-parity violating supersymmetric model}, \emph{Phys. Rev.} {\bf D83},
  \penalty0 115019  (2011).
\newblock \doi{10.1103/PhysRevD.83.115019}.

\bibitem{Chakraborty:2015bsk}
A.~Chakraborty and S.~Chakraborty, {Probing $(g-2)_{\mu}$ at the LHC in the
  paradigm of $R$-parity violating MSSM}, \emph{Phys. Rev.} {\bf D93}\penalty0
  (7), \penalty0 075035  (2016).
\newblock \doi{10.1103/PhysRevD.93.075035}.

\bibitem{Deshpande:2012rr}
N.~G. Deshpande and A.~Menon, {Hints of R-parity violation in B decays into
  $\tau \nu$}, \emph{JHEP}. {\bf 01}, \penalty0 025  (2013).
\newblock \doi{10.1007/JHEP01(2013)025}.

\bibitem{Biswas:2014gga}
S.~Biswas, D.~Chowdhury, S.~Han, and S.~J. Lee, {Explaining the lepton
  non-universality at the LHCb and CMS within a unified framework},
  \emph{JHEP}. {\bf 02}, \penalty0 142  (2015).
\newblock \doi{10.1007/JHEP02(2015)142}.

\bibitem{Huang:2015vpt}
W.~Huang and Y.-L. Tang, {Flavor anomalies at the LHC and the R-parity
  violating supersymmetric model extended with vectorlike particles},
  \emph{Phys. Rev.} {\bf D92}\penalty0 (9), \penalty0 094015  (2015).
\newblock \doi{10.1103/PhysRevD.92.094015}.

\bibitem{Zhu:2016xdg}
J.~Zhu, H.-M. Gan, R.-M. Wang, Y.-Y. Fan, Q.~Chang, and Y.-G. Xu, {Probing the
  R-parity violating supersymmetric effects in the exclusive $b\to
  c\ell^-\bar{\nu}_\ell$ decays}, \emph{Phys. Rev.} {\bf D93}\penalty0 (9),
  \penalty0 094023  (2016).
\newblock \doi{10.1103/PhysRevD.93.094023}.

\bibitem{Deshpand:2016cpw}
N.~G. Deshpande and X.-G. He, {Consequences of R-parity violating interactions
  for anomalies in $\bar B\to D^{(*)} \tau \bar \nu$ and $b\to s \mu^+\mu^-$},
  \emph{Eur. Phys. J.} {\bf C77}\penalty0 (2), \penalty0 134  (2017).
\newblock \doi{10.1140/epjc/s10052-017-4707-y}.

\bibitem{Altmannshofer:2017poe}
W.~Altmannshofer, P.~B. Dev, and A.~Soni, {$R_{D^{(*)}}$ anomaly: A possible
  hint for natural supersymmetry with $R$-parity violation}, \emph{Phys. Rev.}
  {\bf D96}\penalty0 (9), \penalty0 095010  (2017).
\newblock \doi{10.1103/PhysRevD.96.095010}.

\bibitem{Das:2017kfo}
D.~Das, C.~Hati, G.~Kumar, and N.~Mahajan, {Scrutinizing $R$-parity violating
  interactions in light of $R_{K^{(\ast)}}$ data}, \emph{Phys. Rev.} {\bf
  D96}\penalty0 (9), \penalty0 095033  (2017).
\newblock \doi{10.1103/PhysRevD.96.095033}.

\bibitem{Earl:2018snx}
K.~Earl and T.~Grégoire, {Contributions to ${b \rightarrow s \ell \ell}$
  Anomalies from ${R}$-Parity Violating Interactions}, \emph{JHEP}. {\bf 08},
  \penalty0 201  (2018).
\newblock \doi{10.1007/JHEP08(2018)201}.

\bibitem{Sheng:2018ylg}
J.-H. Sheng, J.-J. Song, R.-M. Wang, and Y.-D. Yang, {The lepton flavor
  violating exclusive $b\to s\ell\ell$ decays in SUSY without R-parity},
  \emph{Nucl. Phys.} {\bf B930}, \penalty0 69--90  (2018).
\newblock \doi{10.1016/j.nuclphysb.2018.02.019}.

\bibitem{Trifinopoulos:2018rna}
S.~Trifinopoulos, {Revisiting R-parity violating interactions as an explanation
  of the B-physics anomalies}, \emph{Eur. Phys. J.} {\bf C78}\penalty0 (10),
  \penalty0 803  (2018).
\newblock \doi{10.1140/epjc/s10052-018-6280-4}.

\bibitem{Hu:2018lmk}
Q.-Y. Hu, X.-Q. Li, Y.~Muramatsu, and Y.-D. Yang, {R-parity violating solutions
  to the $R_{D^{(\ast)}}$ anomaly and their GUT-scale unifications},
  \emph{Phys. Rev.} {\bf D99}\penalty0 (1), \penalty0 015008  (2019).
\newblock \doi{10.1103/PhysRevD.99.015008}.

\bibitem{Collins:2018jpg}
J.~H. Collins, P.~S.~B. Dev, and Y.~Sui, {R-parity Violating Supersymmetric
  Explanation of the Anomalous Events at ANITA}, \emph{Phys. Rev.} {\bf
  D99}\penalty0 (4), \penalty0 043009  (2019).
\newblock \doi{10.1103/PhysRevD.99.043009}.

\bibitem{Carena:1998gd}
M.~Carena, D.~Choudhury, S.~Lola, and C.~Quigg, {Manifestations of R-parity
  violation in ultrahigh-energy neutrino interactions}, \emph{Phys. Rev.} {\bf
  D58}, \penalty0 095003  (1998).
\newblock \doi{10.1103/PhysRevD.58.095003}.

\bibitem{Dev:2016uxj}
P.~S.~B. Dev, D.~K. Ghosh, and W.~Rodejohann, {R-parity Violating Supersymmetry
  at IceCube}, \emph{Phys. Lett.} {\bf B762}, \penalty0 116--123  (2016).
\newblock \doi{10.1016/j.physletb.2016.08.066}.

\bibitem{Albuquerque:2003mi}
I.~Albuquerque, G.~Burdman, and Z.~Chacko, {Neutrino telescopes as a direct
  probe of supersymmetry breaking}, \emph{Phys. Rev. Lett.} {\bf 92}, \penalty0
  221802  (2004).
\newblock \doi{10.1103/PhysRevLett.92.221802}.

\bibitem{Ahlers:2006pf}
M.~Ahlers, J.~Kersten, and A.~Ringwald, {Long-lived staus at neutrino
  telescopes}, \emph{JCAP}. {\bf 0607}, \penalty0 005  (2006).
\newblock \doi{10.1088/1475-7516/2006/07/005}.

\bibitem{Albuquerque:2006am}
I.~F.~M. Albuquerque, G.~Burdman, and Z.~Chacko, {Direct detection of
  supersymmetric particles in neutrino telescopes}, \emph{Phys. Rev.} {\bf
  D75}, \penalty0 035006  (2007).
\newblock \doi{10.1103/PhysRevD.75.035006}.

\bibitem{Ando:2007ds}
S.~Ando, J.~F. Beacom, S.~Profumo, and D.~Rainwater, {Probing new physics with
  long-lived charged particles produced by atmospheric and astrophysical
  neutrinos}, \emph{JCAP}. {\bf 0804}, \penalty0 029  (2008).
\newblock \doi{10.1088/1475-7516/2008/04/029}.

\bibitem{Canadas:2008ey}
B.~Canadas, D.~G. Cerdeno, C.~Munoz, and S.~Panda, {Stau detection at neutrino
  telescopes in scenarios with supersymmetric dark matter}, \emph{JCAP}. {\bf
  0904}, \penalty0 028  (2009).
\newblock \doi{10.1088/1475-7516/2009/04/028}.

\bibitem{Meade:2009mu}
P.~Meade, S.~Nussinov, M.~Papucci, and T.~Volansky, {Searches for Long Lived
  Neutral Particles}, \emph{JHEP}. {\bf 06}, \penalty0 029  (2010).
\newblock \doi{10.1007/JHEP06(2010)029}.

\bibitem{Connolly:2018ewv}
A.~Connolly, P.~Allison, and O.~Banerjee, {On ANITA's sensitivity to
  long-lived, charged massive particles}  (2018).
\newblock 1807.08892.

\bibitem{Glashow:1960zz}
S.~L. Glashow, {Resonant Scattering of Antineutrinos}, \emph{Phys. Rev.} {\bf
  118}, \penalty0 316--317  (1960).
\newblock \doi{10.1103/PhysRev.118.316}.

\bibitem{Aartsen:2017mau}
M.~G. Aartsen et~al., {The IceCube Neutrino Observatory - Contributions to ICRC
  2017 Part II: Properties of the Atmospheric and Astrophysical Neutrino Flux}
  (2017).
\newblock 1710.01191.

\bibitem{Fayet:1974pd}
P.~Fayet, {Supergauge Invariant Extension of the Higgs Mechanism and a Model
  for the electron and Its Neutrino}, \emph{Nucl. Phys.} {\bf B90}, \penalty0
  104--124  (1975).
\newblock \doi{10.1016/0550-3213(75)90636-7}.

\bibitem{Farrar:1978xj}
G.~R. Farrar and P.~Fayet, {Phenomenology of the Production, Decay, and
  Detection of New Hadronic States Associated with Supersymmetry}, \emph{Phys.
  Lett.} {\bf 76B}, \penalty0 575--579  (1978).
\newblock \doi{10.1016/0370-2693(78)90858-4}.

\bibitem{DeRujula:1989fe}
A.~De~Rujula, S.~L. Glashow, and U.~Sarid, {CHARGED DARK MATTER}, \emph{Nucl.
  Phys.} {\bf B333}, \penalty0 173--194  (1990).
\newblock \doi{10.1016/0550-3213(90)90227-5}.

\bibitem{Dimopoulos:1989hk}
S.~Dimopoulos, D.~Eichler, R.~Esmailzadeh, and G.~D. Starkman, {Getting a
  Charge Out of Dark Matter}, \emph{Phys. Rev.} {\bf D41}, \penalty0 2388
  (1990).
\newblock \doi{10.1103/PhysRevD.41.2388}.

\bibitem{Pospelov:2006sc}
M.~Pospelov, {Particle physics catalysis of thermal Big Bang Nucleosynthesis},
  \emph{Phys. Rev. Lett.} {\bf 98}, \penalty0 231301  (2007).
\newblock \doi{10.1103/PhysRevLett.98.231301}.

\bibitem{Jungman:1995df}
G.~Jungman, M.~Kamionkowski, and K.~Griest, {Supersymmetric dark matter},
  \emph{Phys. Rept.} {\bf 267}, \penalty0 195--373  (1996).
\newblock \doi{10.1016/0370-1573(95)00058-5}.

\bibitem{Nath:2006ut}
P.~Nath and P.~Fileviez~Perez, {Proton stability in grand unified theories, in
  strings and in branes}, \emph{Phys. Rept.} {\bf 441}, \penalty0 191--317
  (2007).
\newblock \doi{10.1016/j.physrep.2007.02.010}.

\bibitem{Takayama:2000uz}
F.~Takayama and M.~Yamaguchi, {Gravitino dark matter without R-parity},
  \emph{Phys. Lett.} {\bf B485}, \penalty0 388--392  (2000).
\newblock \doi{10.1016/S0370-2693(00)00726-7}.

\bibitem{LopezFogliani:2005yw}
D.~E. Lopez-Fogliani and C.~Munoz, {Proposal for a Supersymmetric Standard
  Model}, \emph{Phys. Rev. Lett.} {\bf 97}, \penalty0 041801  (2006).
\newblock \doi{10.1103/PhysRevLett.97.041801}.

\bibitem{Buchmuller:2007ui}
W.~Buchmuller, L.~Covi, K.~Hamaguchi, A.~Ibarra, and T.~Yanagida, {Gravitino
  Dark Matter in R-Parity Breaking Vacua}, \emph{JHEP}. {\bf 03}, \penalty0 037
   (2007).
\newblock \doi{10.1088/1126-6708/2007/03/037}.

\bibitem{Monteux:2014tia}
A.~Monteux, E.~Carlson, and J.~Cornell, {Gravitino Dark Matter and Flavor
  Symmetries}, \emph{JHEP}. {\bf 08}, \penalty0 047  (2014).
\newblock \doi{10.1007/JHEP08(2014)047}.

\bibitem{DiLuzio:2013ysa}
L.~Di~Luzio, M.~Nardecchia, and A.~Romanino, {Framework for baryonic R-parity
  violation in grand unified theories}, \emph{Phys. Rev.} {\bf D88}\penalty0
  (11), \penalty0 115008  (2013).
\newblock \doi{10.1103/PhysRevD.88.115008}.

\bibitem{Bajc:2015zja}
B.~Bajc and L.~Di~Luzio, {R-parity violation in SU(5)}, \emph{JHEP}. {\bf 07},
  \penalty0 123  (2015).
\newblock \doi{10.1007/JHEP07(2015)123}.

\bibitem{Martin:1992mq}
S.~P. Martin, {Some simple criteria for gauged R-parity}, \emph{Phys. Rev.}
  {\bf D46}, \penalty0 R2769--R2772  (1992).
\newblock \doi{10.1103/PhysRevD.46.R2769}.

\bibitem{Mohapatra:2015fua}
R.~N. Mohapatra, {Supersymmetry and R-parity: an Overview}, \emph{Phys.
  Scripta}. {\bf 90}, \penalty0 088004  (2015).
\newblock \doi{10.1088/0031-8949/90/8/088004}.

\bibitem{Aulakh:1982yn}
C.~S. Aulakh and R.~N. Mohapatra, {Neutrino as the Supersymmetric Partner of
  the Majoron}, \emph{Phys. Lett.} {\bf 119B}, \penalty0 136--140  (1982).
\newblock \doi{10.1016/0370-2693(82)90262-3}.

\bibitem{Robinett:1987ym}
R.~W. Robinett, {The Production of Leptoquarks in Ultrahigh-energy Neutrino
  Interactions}, \emph{Phys. Rev.} {\bf D37}, \penalty0 84--88  (1988).
\newblock \doi{10.1103/PhysRevD.37.84}.

\bibitem{Doncheski:1997it}
M.~A. Doncheski and R.~W. Robinett, {Leptoquark production in ultrahigh-energy
  neutrino interactions revisited}, \emph{Phys. Rev.} {\bf D56}, \penalty0
  7412--7415  (1997).
\newblock \doi{10.1103/PhysRevD.56.7412}.

\bibitem{Anchordoqui:2006wc}
L.~A. Anchordoqui, C.~A. Garcia~Canal, H.~Goldberg, D.~Gomez~Dumm, and
  F.~Halzen, {Probing leptoquark production at IceCube}, \emph{Phys. Rev.} {\bf
  D74}, \penalty0 125021  (2006).
\newblock \doi{10.1103/PhysRevD.74.125021}.

\bibitem{Alikhanov:2013fda}
I.~Alikhanov, {Do leptoquarks manifest themselves in ultra-high energy neutrino
  interactions?}, \emph{JHEP}. {\bf 07}, \penalty0 093  (2013).
\newblock \doi{10.1007/JHEP07(2013)093}.

\bibitem{Barger:2013pla}
V.~Barger and W.-Y. Keung, {Superheavy Particle Origin of IceCube PeV Neutrino
  Events}, \emph{Phys. Lett.} {\bf B727}, \penalty0 190--193  (2013).
\newblock \doi{10.1016/j.physletb.2013.10.021}.

\bibitem{Dutta:2015dka}
B.~Dutta, Y.~Gao, T.~Li, C.~Rott, and L.~E. Strigari, {Leptoquark implication
  from the CMS and IceCube experiments}, \emph{Phys. Rev.} {\bf D91}, \penalty0
  125015  (2015).
\newblock \doi{10.1103/PhysRevD.91.125015}.

\bibitem{Dey:2015eaa}
U.~K. Dey and S.~Mohanty, {Constraints on Leptoquark Models from IceCube Data},
  \emph{JHEP}. {\bf 04}, \penalty0 187  (2016).
\newblock \doi{10.1007/JHEP04(2016)187}.

\bibitem{Dey:2017ede}
U.~K. Dey, D.~Kar, M.~Mitra, M.~Spannowsky, and A.~C. Vincent, {Searching for
  Leptoquarks at IceCube and the LHC}, \emph{Phys. Rev.} {\bf D98}\penalty0
  (3), \penalty0 035014  (2018).
\newblock \doi{10.1103/PhysRevD.98.035014}.

\bibitem{Becirevic:2018uab}
D.~Becirevic, B.~Panes, O.~Sumensari, and R.~Zukanovich~Funchal, {Seeking
  leptoquarks in IceCube}, \emph{JHEP}. {\bf 06}, \penalty0 032  (2018).
\newblock \doi{10.1007/JHEP06(2018)032}.

\bibitem{Ibanez:1991hv}
L.~E. Ibanez and G.~G. Ross, {Discrete gauge symmetry anomalies}, \emph{Phys.
  Lett.} {\bf B260}, \penalty0 291--295  (1991).
\newblock \doi{10.1016/0370-2693(91)91614-2}.

\bibitem{Smirnov:1996bg}
A.~{\relax Yu}. Smirnov and F.~Vissani, {Upper bound on all products of
  R-parity violating couplings lambda-prime and lambda-prime-prime from proton
  decay}, \emph{Phys. Lett.} {\bf B380}, \penalty0 317--323  (1996).
\newblock \doi{10.1016/0370-2693(96)00495-9}.

\bibitem{Hirsch:2007kx}
M.~Hirsch, D.~P. Roy, and J.~W.~F. Valle, {Probing a Supersymmetric Model for
  Neutrino Masses at Ultrahigh Energy Neutrino Telescopes}, \emph{Phys. Lett.}
  {\bf B662}, \penalty0 185--189  (2008).
\newblock \doi{10.1016/j.physletb.2008.02.065}.

\bibitem{Butterworth:1992tc}
J.~Butterworth and H.~K. Dreiner, {R-parity violation at HERA}, \emph{Nucl.
  Phys.} {\bf B397}, \penalty0 3--34  (1993).
\newblock \doi{10.1016/0550-3213(93)90334-L}.

\bibitem{Allanach:1999ic}
B.~C. Allanach, A.~Dedes, and H.~K. Dreiner, {Bounds on R-parity violating
  couplings at the weak scale and at the GUT scale}, \emph{Phys. Rev.} {\bf
  D60}, \penalty0 075014  (1999).
\newblock \doi{10.1103/PhysRevD.60.075014}.

\bibitem{Kao:2009fg}
Y.~Kao and T.~Takeuchi, {Single-Coupling Bounds on R-parity violating
  Supersymmetry, an update}  (2009).
\newblock 0910.4980.

\bibitem{Dreiner:2012mx}
H.~K. Dreiner, K.~Nickel, F.~Staub, and A.~Vicente, {New bounds on trilinear
  R-parity violation from lepton flavor violating observables}, \emph{Phys.
  Rev.} {\bf D86}, \penalty0 015003  (2012).
\newblock \doi{10.1103/PhysRevD.86.015003}.

\bibitem{Domingo:2018qfg}
F.~Domingo, H.~K. Dreiner, J.~S. Kim, M.~E. Krauss, M.~Lozano, and Z.~S. Wang,
  {Updating Bounds on $R$-Parity Violating Supersymmetry from Meson Oscillation
  Data}  (2018).
\newblock 1810.08228.

\bibitem{Gandhi:1995tf}
R.~Gandhi, C.~Quigg, M.~H. Reno, and I.~Sarcevic, {Ultrahigh-energy neutrino
  interactions}, \emph{Astropart. Phys.} {\bf 5}, \penalty0 81--110  (1996).
\newblock \doi{10.1016/0927-6505(96)00008-4}.

\bibitem{CooperSarkar:2011pa}
A.~Cooper-Sarkar, P.~Mertsch, and S.~Sarkar, {The high energy neutrino
  cross-section in the Standard Model and its uncertainty}, \emph{JHEP}. {\bf
  08}, \penalty0 042  (2011).
\newblock \doi{10.1007/JHEP08(2011)042}.

\bibitem{Abbiendi:2003rn}
G.~Abbiendi et~al., {Search for R parity violating decays of scalar fermions at
  LEP}, \emph{Eur. Phys. J.} {\bf C33}, \penalty0 149--172  (2004).
\newblock \doi{10.1140/epjc/s2004-01596-8}.

\bibitem{Abdallah:2003xc}
J.~Abdallah et~al., {Search for supersymmetric particles assuming R-parity
  nonconservation in e+ e- collisions at s**(1/2) = 192-GeV to 208-GeV},
  \emph{Eur. Phys. J.} {\bf C36}\penalty0 (1), \penalty0 1--23  (2004).
\newblock \doi{10.1140/epjc/s2004-01881-6, 10.1140/epjc/s2004-01976-0}.
\newblock [Erratum: Eur. Phys. J.C37,no.1,129(2004)].

\bibitem{Laha:2013eev}
R.~Laha, J.~F. Beacom, B.~Dasgupta, S.~Horiuchi, and K.~Murase, {Demystifying
  the PeV Cascades in IceCube: Less (Energy) is More (Events)}, \emph{Phys.
  Rev.} {\bf D88}, \penalty0 043009  (2013).
\newblock \doi{10.1103/PhysRevD.88.043009}.

\bibitem{Chen:2013dza}
C.-Y. Chen, P.~S.~B. Dev, and A.~Soni, {Standard model explanation of the
  ultrahigh energy neutrino events at IceCube}, \emph{Phys. Rev.} {\bf
  D89}\penalty0 (3), \penalty0 033012  (2014).
\newblock \doi{10.1103/PhysRevD.89.033012}.

\bibitem{Aartsen:2013vja}
M.~G. Aartsen et~al., {Energy Reconstruction Methods in the IceCube Neutrino
  Telescope}, \emph{JINST}. {\bf 9}, \penalty0 P03009  (2014).
\newblock \doi{10.1088/1748-0221/9/03/P03009}.

\bibitem{Aartsen:2013jdh}
M.~G. Aartsen et~al., {Evidence for High-Energy Extraterrestrial Neutrinos at
  the IceCube Detector}, \emph{Science}. {\bf 342}, \penalty0 1242856  (2013).
\newblock \doi{10.1126/science.1242856}.

\bibitem{Vincent:2016nut}
A.~C. Vincent, S.~Palomares-Ruiz, and O.~Mena, {Analysis of the 4-year IceCube
  high-energy starting events}, \emph{Phys. Rev.} {\bf D94}\penalty0 (2),
  \penalty0 023009  (2016).
\newblock \doi{10.1103/PhysRevD.94.023009}.

\bibitem{Palladino:2018evm}
A.~Palladino and W.~Winter, {A multi-component model for observed astrophysical
  neutrinos}, \emph{Astron. Astrophys.} {\bf 615}, \penalty0 A168  (2018).
\newblock \doi{10.3204/PUBDB-2018-01376, 10.1051/0004-6361/201832731}.

\bibitem{Sui:2018bbh}
Y.~Sui and P.~S.~B. Dev, {A Combined Astrophysical and Dark Matter
  Interpretation of the IceCube HESE and Throughgoing Muon Events},
  \emph{JCAP}. {\bf 1807}\penalty0 (07), \penalty0 020  (2018).
\newblock \doi{10.1088/1475-7516/2018/07/020}.

\bibitem{Ball:2012cx}
R.~D. Ball et~al., {Parton distributions with LHC data}, \emph{Nucl. Phys.}
  {\bf B867}, \penalty0 244--289  (2013).
\newblock \doi{10.1016/j.nuclphysb.2012.10.003}.

\bibitem{Collaboration:2010ez}
F.~D. Aaron et~al., {Search for Squarks in R-parity Violating Supersymmetry in
  ep Collisions at HERA}, \emph{Eur. Phys. J.} {\bf C71}, \penalty0 1572
  (2011).
\newblock \doi{10.1140/epjc/s10052-011-1572-y}.

\bibitem{South:2016cmx}
D.~M. South and M.~Turcato, {Review of Searches for Rare Processes and Physics
  Beyond the Standard Model at HERA}, \emph{Eur. Phys. J.} {\bf C76}\penalty0
  (6), \penalty0 336  (2016).
\newblock \doi{10.1140/epjc/s10052-016-4152-3}.

\bibitem{Abbiendi:2001aj}
G.~Abbiendi et~al., {Search for single leptoquark and squark production in
  electron photon scattering at $\sqrt{s_{ee}}$ = 189-GeV at LEP}, \emph{Eur.
  Phys. J.} {\bf C23}, \penalty0 1--11  (2002).
\newblock \doi{10.1007/s100520100859}.

\bibitem{Heister:2002jc}
A.~Heister et~al., {Search for supersymmetric particles with R parity violating
  decays in $e^{+} e^{-}$ collisions at $\sqrt{s}$ up to 209-GeV}, \emph{Eur.
  Phys. J.} {\bf C31}, \penalty0 1--16  (2003).
\newblock \doi{10.1140/epjc/s2003-01311-5}.

\bibitem{Abbott:1999nh}
B.~Abbott et~al., {Search for $R-$parity violating supersymmetry in the
  dielectron channel}, \emph{Phys. Rev. Lett.} {\bf 83}, \penalty0 4476--4481
  (1999).
\newblock \doi{10.1103/PhysRevLett.83.4476}.

\bibitem{Abe:1998gu}
F.~Abe et~al., {Search for $R$-parity violating supersymmetry using like-sign
  dielectrons in $p \bar{p}$ collisions at $\sqrt{s} = 1.8$ TeV}, \emph{Phys.
  Rev. Lett.} {\bf 83}, \penalty0 2133--2138  (1999).
\newblock \doi{10.1103/PhysRevLett.83.2133}.

\bibitem{Aaboud:2019jcc}
M.~Aaboud et~al., {Searches for scalar leptoquarks and differential
  cross-section measurements in dilepton-dijet events in proton-proton
  collisions at a centre-of-mass energy of $\sqrt{s}$ = 13 TeV with the ATLAS
  experiment}  (2019).
\newblock 1902.00377.

\bibitem{Yamanaka:2014nba}
N.~Yamanaka, T.~Sato, and T.~Kubota, {Linear programming analysis of the
  $R$-parity violation within EDM-constraints}, \emph{JHEP}. {\bf 12},
  \penalty0 110  (2014).
\newblock \doi{10.1007/JHEP12(2014)110}.

\bibitem{Ghosh:2001mr}
D.~K. Ghosh, X.-G. He, B.~H.~J. McKellar, and J.-Q. Shi, {Constraining R parity
  violating couplings from B ---> PP decays using QCD improved factorization
  method}, \emph{JHEP}. {\bf 07}, \penalty0 067  (2002).
\newblock \doi{10.1088/1126-6708/2002/07/067}.

\bibitem{Dreiner:2013jta}
H.~K. Dreiner, K.~Nickel, and F.~Staub, {$B^0_{s,d} \to \mu\overline{\mu}$ and
  $B \to X_s\gamma$ in the R-parity violating MSSM}, \emph{Phys. Rev.} {\bf
  D88}\penalty0 (11), \penalty0 115001  (2013).
\newblock \doi{10.1103/PhysRevD.88.115001}.

\bibitem{Agostini:2018tnm}
M.~Agostini et~al., {Improved Limit on Neutrinoless Double-$\beta$ Decay of
  $^{76}$Ge from GERDA Phase II}, \emph{Phys. Rev. Lett.} {\bf 120}\penalty0
  (13), \penalty0 132503  (2018).
\newblock \doi{10.1103/PhysRevLett.120.132503}.

\bibitem{Kou:2018nap}
W.~Altmannshofer et~al., {The Belle II Physics Book}  (2018).
\newblock 1808.10567.

\bibitem{Adrian-Martinez:2016fdl}
S.~Adrian-Martinez et~al., {Letter of intent for KM3NeT 2.0}, \emph{J. Phys.}
  {\bf G43}\penalty0 (8), \penalty0 084001  (2016).
\newblock \doi{10.1088/0954-3899/43/8/084001}.

\bibitem{Aartsen:2014njl}
M.~G. Aartsen et~al., {IceCube-Gen2: A Vision for the Future of Neutrino
  Astronomy in Antarctica}  (2014).
\newblock 1412.5106.

\bibitem{Giudice:1998bp}
G.~F. Giudice and R.~Rattazzi, {Theories with gauge mediated supersymmetry
  breaking}, \emph{Phys. Rept.} {\bf 322}, \penalty0 419--499  (1999).
\newblock \doi{10.1016/S0370-1573(99)00042-3}.

\bibitem{Reno:2005si}
M.~H. Reno, I.~Sarcevic, and S.~Su, {Propagation of supersymmetric charged
  sleptons at high energies}, \emph{Astropart. Phys.} {\bf 24}, \penalty0
  107--115  (2005).
\newblock \doi{10.1016/j.astropartphys.2005.06.002}.

\bibitem{Khachatryan:2016sfv}
V.~Khachatryan et~al., {Search for long-lived charged particles in
  proton-proton collisions at $\sqrt{s} = 13$ TeV}, \emph{Phys. Rev.} {\bf
  D94}\penalty0 (11), \penalty0 112004  (2016).
\newblock \doi{10.1103/PhysRevD.94.112004}.

\bibitem{Gorham:2016zah}
P.~W. Gorham et~al., {Characteristics of Four Upward-pointing Cosmic-ray-like
  Events Observed with ANITA}, \emph{Phys. Rev. Lett.} {\bf 117}\penalty0 (7),
  \penalty0 071101  (2016).
\newblock \doi{10.1103/PhysRevLett.117.071101}.

\bibitem{Gorham:2018ydl}
P.~W. Gorham et~al., {Observation of an Unusual Upward-going Cosmic-ray-like
  Event in the Third Flight of ANITA}, \emph{Phys. Rev. Lett.} {\bf
  121}\penalty0 (16), \penalty0 161102  (2018).
\newblock \doi{10.1103/PhysRevLett.121.161102}.

\bibitem{Fox:2018syq}
D.~B. Fox, S.~Sigurdsson, S.~Shandera, P.~Mészáros, K.~Murase, M.~Mostafá,
  and S.~Coutu, {The ANITA Anomalous Events as Signatures of a Beyond Standard
  Model Particle, and Supporting Observations from IceCube}  (2018).
\newblock 1809.09615.

\bibitem{Romero-Wolf:2018zxt}
A.~Romero-Wolf et~al., {A comprehensive analysis of anomalous ANITA events
  disfavors a diffuse tau-neutrino flux origin}  (2018).
\newblock 1811.07261.

\bibitem{Greisen:1966jv}
K.~Greisen, {End to the cosmic ray spectrum?}, \emph{Phys. Rev. Lett.} {\bf
  16}, \penalty0 748--750  (1966).
\newblock \doi{10.1103/PhysRevLett.16.748}.

\bibitem{Zatsepin:1966jv}
G.~T. Zatsepin and V.~A. Kuzmin, {Upper limit of the spectrum of cosmic rays},
  \emph{JETP Lett.} {\bf 4}, \penalty0 78--80  (1966).
\newblock [Pisma Zh. Eksp. Teor. Fiz.4,114(1966)].

\bibitem{Barnard:2012au}
J.~Barnard, B.~Farmer, T.~Gherghetta, and M.~White, {Natural gauge mediation
  with a bino NLSP at the LHC}, \emph{Phys. Rev. Lett.} {\bf 109}, \penalty0
  241801  (2012).
\newblock \doi{10.1103/PhysRevLett.109.241801}.

\bibitem{Learned:1994wg}
J.~G. Learned and S.~Pakvasa, {Detecting tau-neutrino oscillations at PeV
  energies}, \emph{Astropart. Phys.} {\bf 3}, \penalty0 267--274  (1995).
\newblock \doi{10.1016/0927-6505(94)00043-3}.

\bibitem{Aad:2014iza}
G.~Aad et~al., {Search for supersymmetry in events with four or more leptons in
  $\sqrt{s}$ = 8 TeV pp collisions with the ATLAS detector}, \emph{Phys. Rev.}
  {\bf D90}\penalty0 (5), \penalty0 052001  (2014).
\newblock \doi{10.1103/PhysRevD.90.052001}.

\bibitem{Aaboud:2017vwy}
M.~Aaboud et~al., {Search for squarks and gluinos in final states with jets and
  missing transverse momentum using $\sqrt{s} = 13$ TeV pp collision data with
  the ATLAS detector}, \emph{Phys. Rev.} {\bf D97}\penalty0 (11), \penalty0
  112001  (2018).
\newblock \doi{10.1103/PhysRevD.97.112001}.

\end{thebibliography}

\end{document}